\documentclass[10pt,aps,twocolumn,showpacs,superscriptaddress,floatfix]{revtex4-1}
\usepackage{graphicx,amsmath}
\usepackage{epsfig}
\usepackage{newlfont,bm,dcolumn,longtable}
\usepackage{amssymb,amsfonts,amsmath}
\usepackage{color}
\usepackage{eucal}
\usepackage{natbib}
\usepackage{xspace}
\usepackage{hyperref}

\newcommand{\thav}[1]{\left< #1 \right>}
\newcommand{\smav}[1]{\overline{#1}}

\newcommand{\EAI}{EAI\xspace}

\begin{document}

\title{
The cumulative overlap distribution function in realistic spin glasses
}

\author{A. Billoire}
\affiliation{Institut de physique th\'eorique, CEA Saclay and CNRS, 91191
  Gif-sur-Yvette, France} 
\author{A. Maiorano}
\affiliation{Dipartimento di Fisica,
  Sapienza Universit\`a di Roma, P. A. Moro 2, 00185 Roma, Italy}
\affiliation{Instituto de Biocomputaci\'on y F\'{\i}sica de Sistemas Complejos (BIFI)
  50018 Zaragoza, Spain}
\author{E. Marinari}
\affiliation{Dipartimento di Fisica, IPCF-CNR and INFN,
  Sapienza Universit\`a di Roma, P. A. Moro 2, 00185 Roma, Italy} 
\author{V. Martin-Mayor}
\affiliation{Departamento de F\'{\i}sica Te\'orica, Facultad de
  Ciencias F\'{\i}sicas,\\
Universidad Complutense de Madrid, 28040 Madrid, Spain}
\affiliation{Instituto de Biocomputaci\'on y F\'{\i}sica de Sistemas Complejos (BIFI)
  50018 Zaragoza, Spain}
\author{D. Yllanes}
\affiliation{Dipartimento di Fisica,
  Sapienza Universit\`a di Roma, P. A. Moro 2, 00185 Roma, Italy}
\affiliation{Instituto de Biocomputaci\'on y F\'{\i}sica de Sistemas Complejos (BIFI)
  50018 Zaragoza, Spain}

\date{\today}

\begin{abstract}
We use a sample-dependent analysis, based on medians and quantiles, to
analyze the behavior of the overlap probability distribution of the
Sherrington-Kirkpatrick and 3D Edwards-Anderson models of Ising spin
glasses.  We find that this approach is an effective tool to
distinguish between RSB-like and droplet-like behavior of the
spin-glass phase. Our results are in agreement with a RSB-like
behavior for the 3D Edwards-Anderson model.
\end{abstract}

\pacs{75.50.Lk,64.70.Pf,75.10.Hk}

\maketitle

\section{INTRODUCTION}
\label{sec:intro}

The Edwards-Anderson Ising spin glass~\cite{edwards:75} (\EAI) is a
paradigmatic model for disordered magnets. The physics of its fully
connected counterpart, the Sherrington-Kirkpatrick model
(SK) \cite{sherrington:75}, is well
understood~\cite{parisi:80,talagrand:06,mezard:87}. The SK model has
striking features at temperatures below the spin-glass transition
temperature, like replica symmetry breaking (RSB), an ultrametric
organization of the states, and a non-trivial functional order
parameter. The situation is less clear for finite-dimensional spin
glasses. Two conflicting approaches to describe the nature of their
spin-glass phase have gained polarized consensus in the last decades. On
one side, the scaling picture (or equivalently the droplet
model~\cite{fisher:87,*fisher:86,*fisher:88b,bray:87}) describes the
equilibrium properties at low temperatures in terms of a single
thermodynamic state (actually, due to the global spin reversal
symmetry, one pair of states). On the other side the RSB picture, a
mean-field-like description based on the solution of the SK model,
predicts the existence of infinitely many pure states contributing to
the thermodynamic limit.  We stress that the \emph{one or many} states
question is the crucial one~\footnote{An intermediate ``chaotic
  pairs'' picture has been proposed by Newman and
  Stein~\cite{newman:92,*newman:96b}, where many states exist but only
  one, which would depend chaotically on $L$, is manifest in a finite volume.}, whereas the ultrametric
structure of phase space, after recent theoretical
results~\cite{aizenman:98,ghirlanda:98,panchenko:13}, is expected to
hold in many finite-dimensional spin-glass models (at least
trivially), and has been confirmed by numerical experiments
either directly~\cite{contucci:07,janus:11,maiorano:14}, or by
inspection of the overlap equivalence
property~\cite{parisi:00,janus:10}.

Interestingly, the ``one or many'' states question can be cast as well
as a problem about self-averageness. According to the droplet picture,
sample-to-sample fluctuations should fade away in that limit. On the
other hand, a most surprising feature of the mean-field solution is
that these fluctuations survive the thermodynamic limit, and are
substantial~\cite{mezard:84,*mezard:84b,parisi:93}: macroscopic
observable quantities can take different values in different infinite-volume
samples. It is important to note that this disagreement among the two
theories concerns thermal equilibrium, and is thus mostly relevant to
analytic and numerical computations. Experimentally, the question
could be addressed by studying spatial regions as small as the
spin-glass coherence length (that has been estimated to be of the
order of $100$ lattice spacings close to the critical temperature
$T\sim T_\mathrm{c}$~\cite{joh:99}, and smaller for lower and higher
temperatures). This is a difficult approach that is only at its
birth~\cite{oukris:10,komatsu:11}.

The lack of self-averageness, with an emphasis on quantities that have
not been averaged over the quenched
disorder~\cite{janus:11,billoire:11} and especially on the effect of
rare non-typical samples (using either a
numerical~\cite{billoire:14,fernandez:13,janus:14c,monthus:13} or a
theoretical~\cite{rizzo:14} approach), has recently become a topic of
interest. It seems the most promising approach to the study of
temperature chaos~\cite{mckay:82,bray:87b,banavar:87} and the possibly
related rejuvenation and memory effects~\cite{jonason:98}. Therefore,
it is hardly surprising that recent proposals have tried to deal with
the ``one or many'' states controversy by studying sample-to-sample
fluctuations and their system-size
dependency~\cite{yucesoy:12,middleton:13}.  The original approach of
Ref.~\onlinecite{yucesoy:12} has the drawback of not being directly
sensitive to the statistical weight of the states, and a further
improvement to this analysis scheme that has been introduced
recently~\cite{middleton:13} can be of help.

Here, we further refine the approach of
Refs.~\onlinecite{yucesoy:12,middleton:13}, and we compare its
predictions to the different theoretical expectations.  On the one
side RSB predicts large sample-to-sample fluctuations (the probability
density function is barely normalizable), which call for special care
in the data analysis. On the other hand, following
Ref.~\onlinecite{middleton:13}, we employ toy models in order to get
droplet-model like predictions. Both theoretical expectations are
tested against the results of our numerical analysis both for the
three-dimensional \EAI model~\cite{janus:10}, and for the mean-field
SK model. We find that the SK and \EAI models behave much in the same
way (including the finite-size and finite-statistics effects). The
droplets picture is thus disfavored from our analysis, at least within
the range of system sizes and temperatures that one can equilibrate
using the special-purpose Janus computer\cite{janus:09,janus:12b}.

The remaining part of this work is organized as follows. In
Sec.~\ref{sec:Model-and-definitinions} we introduce the model and
provide the crucial definitions. In Sec.~\ref{sec:STATS} we introduce
the quantile statistics that we use to analyze our numerical data, and
we discuss the theoretical expectations. In Sec.~\ref{sec:MC} we
compare our numerical findings with the mean-field predictions.  In
order to test the hypothesis of a droplet spin-glass phase and to
examine the importance of a many-states picture, in
Sec.~\ref{sec:TOYS} we introduce and discuss a few toy models.
Sec.~\ref{sec:CONC} contains an overall discussion and our
conclusions.

\section{MODEL AND  MAIN DEFINITIONS}
\label{sec:Model-and-definitinions}

The Edwards-Anderson model is defined by a nearest-neighbor
Hamiltonian $H$. $H$ is a function of
a set of quenched random coupling constants
$\{J_{ij}\}$ (a specific realization of the random coupling constants is
called a 
disorder sample) and of a set of Ising spin variables $\{\sigma_{i}\}$
defined on the vertices of a (hyper-)cubic lattice: 
\begin{equation}
  \label{eq:EA}
  H \equiv - \sum_{\langle i,j \rangle} J_{ij} \sigma_{i} \sigma_{j}\;, 
\end{equation}
where the summation extends over all pairs of nearest-neighboring sites.  The
couplings $\{J_{ij}\}$ are independent and identically distributed (i.i.d.) random
variables with zero mean and unit variance, usually standard normal or, as in
our numerical experiments, binary ($\pm 1$) distributed.  In the SK model,
every spin interacts with all other spins, and the variance of the $\{J_{ij}\}$
distribution is inversely proportional to the total number of spins.

It is an established fact, with both experimental~\cite{gunnarsson:91}
and numerical evidence~\cite{palassini:99,ballesteros:00}, that in
three spatial dimensions the \EAI model undergoes a second-order phase
transition at a finite transition temperature, from a paramagnetic
high-temperature state to a low-temperature spin-glass state (with no
magnetic long-range order).  The \emph{overlap} between two independent
equilibrium spin configurations in the same disorder 
sample (two real replicas) is the order parameter of the model:
\begin{equation}
\label{eq:q}
q=\frac{1}{N}\sum_i\sigma_i^a\sigma_j^b\ \;, 
\end{equation} 
where $N$ is the total
number of spins (In the $3D$ EAI case $N=L^D$, where $D$ is the spatial dimension and  
$L$ is the linear size
of the lattice).  The overlap is a random variable whose probability
density $P_J(q)$ depends on the disorder realization. The
overlap distribution is the average over all disorder
realizations of $P_J(q)$: 
\begin{eqnarray}
  \label{eq:Pq}
  P(q) & = & \smav{P_J(q)}\ \;, \\ 
  P_J(q) & = & \thav{\delta\left(
    q - \frac{1}{N}\sum_i\sigma_i^a\sigma_j^b
  \right)}\ \;, 
\end{eqnarray} 
where we adopt the usual notation $\thav{\cdots}$ for thermal averages
in a single disorder sample and $\smav{\cdots}$ for the average over
different samples.

The droplet and RSB pictures offer dramatically different
qualitative predictions for the shape of $P(q)$ in the thermodynamic
limit.  In the droplet scenario, for large system sizes, a single delta
function and its global-inversion symmetric image (both smeared by finite-size
effects) dominate the overlap distribution; the location of this
delta function defines the \emph{Edwards-Anderson order parameter}
$q_\text{EA}=\smav{\thav{\sigma_i}^2}$. At high temperature, $q_\text{EA}$ is
null; below the transition temperature the spins freeze in disordered
(sample-dependent) orientations and the overlap distribution is a
symmetric pair of (smeared) delta functions at $\pm q_\text{EA}$.  In the
RSB scenario a continuous distribution is present between the two
symmetric delta peaks at $q=\pm q_\text{EA}$, due to the presence of
infinitely many states in the thermodynamic limit.

Since the predictions of the droplet and RSB pictures for the behavior
of $P(q\approx 0)$ are so different, precise numerical measurements of
the quantities in Eq.~(\ref{eq:Pq}) could in principle give a
clear-cut distinction between the two pictures. Numerical simulations
are, however, always performed on finite systems and accordingly an
extrapolation to the infinite-volume limit is needed. This
extrapolation is, however, not straightforward, and the question of
the large-volume limit of $P(q\approx 0)$ data has led to contrasting
interpretations in the
literature~\cite{moore:98,*katzgraber:01,*palassini:01,janus:10}.

In the droplet model, compact excitations of linear size $\ell$ have
probability $\ell^{-\theta}$, where $\theta$ is a positive exponent,
and consequently the probability of having small overlap values,
dominated by very large-scale excitations, is vanishing in the
thermodynamic limit as $L^{-\theta}$, with $\theta\sim0.2$ in three
dimensions. This is a very small value, and since the simulations are
performed on small systems (we will present data for $D=3$ systems
with values of $L$ going up to $32$, but many equilibrium numerical
simulations in the literature are limited to $L=12$ or even less) it
is a challenge to distinguish unambiguously between an
$L^{-\theta}$ and a constant limiting behavior of the data.

This has led to a recent shift of attention toward the study of the
whole distribution (with respect to disorder) of the non-averaged
$P_J(q)$, in the quest for a measurable quantity with unmistakably
different finite-volume behavior for the two pictures.  This is part
of a general recent interest on non-disorder-averaged
quantities~\cite{janus:11,billoire:11} and especially on the effect of
rare non-typical 
samples~\cite{fernandez:13,monthus:13,billoire:14,janus:14c,rizzo:14}.

In particular, Ref.~\onlinecite{yucesoy:12} analyzes the probability
$\Delta(\kappa,q_0)$ of finding in $P_J(q)$, for $q<q_0$, a peak
higher than some value $\kappa$.  In the RSB picture this probability
goes to one in the infinite-volume limit for all values
$q_0<q_\text{EA}$.  In the droplet picture, some peaks may exist in
$P_J(q)$ below $q_\text{EA}$, but their effect disappears as $N$
grows, and $\Delta(\kappa,q_0)$ goes to zero in the $N\to\infty$
limit.  The results of Ref.~\onlinecite{yucesoy:12} seem to suggest that
for the SK model this quantity does grow as $N$ grows, but reaches a
plateau for the $3D$ \EAI model.  It was later shown \cite{billoire:13},
however, by using larger systems ($N\leq 32^3$
instead of $12^3$), that $\Delta(\kappa,q_0)$ does grow with $N$ for
the $3D$ \EAI model also. The slower growth that one has in the \EAI model as
compared to the SK model can be explained by the simple assumption that the
peaks for all values of $q$ grow at the same rate, taking into account the
known scaling of the $q_\text{EA}$ peak in both models. Note that a drawback of
the method of Ref.~\onlinecite{yucesoy:12} is that $\Delta(\kappa,q_0)$ is not
directly sensitive to the peak weight, which is what matters here, but to its
height.

In a recent paper~\cite{middleton:13} it was noticed that for both the
zero temperature $2D$ \EAI model with binary distributed couplings, and
for a variant of  the toy droplet model of Ref.~\onlinecite{hatano:02},  the
decrease with $L$ of the average $P(q)$ for small $q$ was very
slow, but the decrease of the median (over the disorder) was much faster.
In fact, this median was compatible with zero for the larger systems
studied in a whole interval of low $q$ values. 
Therefore, the author advocated the study of this median as a
silver bullet to distinguish a droplet-like behavior from a RSB-like
behavior using numerical data.

In this paper we will develop this approach by studying the quantiles
of the sample-dependent cumulative overlap distribution function
\begin{equation}
  \label{eq:XJq}
  X_J(q)\equiv\int_{-q}^qP_J(q^\prime)dq^\prime\;,
\end{equation}
whereas most numerical studies in the past focused on the disorder average
\begin{equation}
  \label{eq:Xq}
  X(q)\equiv \smav{X_J(q)}\;,
\end{equation}
and on low-order moments~\cite{mezard:84,*mezard:84b,janus:11}. 
$X(q)$ is the functional order parameter in RSB
theory~\cite{parisi:80}. In a droplet picture, the average cumulative
overlap distribution is expected to tend to a Heaviside step function
$\theta(q-q_\text{EA})$ for large system sizes.

\section{STATISTICS OF THE CUMULATIVE OVERLAP DISTRIBUTION IN THE MEAN-FIELD PICTURE}
\label{sec:STATS}

In this section we introduce the statistical quantities that we will
study in the following, and we review the mean-field predictions for
their behavior (the droplet picture does not imply any quantitative
prediction about the scaling behavior, and we will address it through
toy models, see Sec.~\ref{sec:TOYS}).

The RSB mean-field analysis of the SK model offers precise
predictions~\cite{mezard:84,*mezard:84b,parisi:93} (valid in the
thermodynamic limit) on the statistics of the random variable $X_J(q)$
defined in Eq.~(\ref{eq:XJq}).  The probability density $\mathbb{P}(X_J=s)$, 
sometimes denoted by $\Pi(s)$ in the literature~\cite{mezard:87},
diverges at the origin as a power law, with an exponent equal to the
average integrated overlap $X(q)$ minus one ($\mathbb{P}(s)$ has rather
complex properties, with an infinite number of singular points,~\cite{parisi:93,derrida:87} but the singularity at the origin
is the strongest). For small $s$ one has that
\begin{equation}
  \label{eq:pow}
  \mathbb{P}(X_J(q)=s) \propto s^{X(q)-1}\;.  
\end{equation} 
In the whole interval $s \in [0,1]$ the density $\mathbb{P}(X_J(q)=s)$
depends on $T$ and $q$ through $X(q)$ only.  According
to~\eqref{eq:pow}, for small $s$ the cumulative probability
$\mathbb{P}_C(s)\equiv\int_0^s \mathbb{P}(s') ds'$, behaves as
\begin{equation}
  \label{eq:powCum}
  \mathbb{P}_C(s) \sim  s^{X(q)}\;.
\end{equation} 
We denote by $I_p(q)$ the first $1/p$-quantile of the integrated
overlaps at fixed $q$, i.e., the value of $X_J(q)$ for which the
probability that $X_J<I_p$ is smaller or equal than $p$, and, at the
same time, the probability that $X_J>I_p$ is smaller or equal than
$1-p$~\footnote{This definition accounts for the presence of a finite weight in
a single point, as one has, for example, in the Parisi mean field
theory}.  We denote by $I(q)\equiv I_{1/2}(q)$ the \emph{median} of
$X_J(q)$; we drop the subscript $p$ when $p=1/2$. For values of $s$
where the cumulative function $\mathbb{P}_C(s)$ is continuous $I_p$ is the
value such that $\mathbb{P}_C(I_p) = p$. Because of that, for small values of
the argument, where (\ref{eq:powCum}) is valid, one has that
\begin{equation}
\label{eq:fpq} 
f_p(q) \equiv  I_p(q)^{X(q)} \sim p\quad \mbox{ for }I_p(q) \to 0\;.
\end{equation} 
This implies that $I_p(q)$ is exponentially small as $X(q)\to 0$, with
$\ln(I_p(q))\sim -\ln(1/p)/X(q)$.  It is equivalent to consider $q$
going to zero, since $q\sim X(q)/2P(0)$.  The distribution of $X_J(q)$
is very skewed for small values of $q$. The most probable value is
zero, and a majority of samples have values of $X_J(q)$ much smaller
than the average.  The average receives non-negligible contributions
from a minority of samples only.

If Eq.~(\ref{eq:fpq}) were exact for all values of $q$, $f_p(q)$ would
be constant (namely $f_p(q)=p$). Eq.~(\ref{eq:fpq}) is, however, only
valid in the small-$q$ region, and it is self-consistent only in this
region as it predicts that the average value of $X_J(q)$ is
\begin{equation}
\label{eq:self}
  X(q) = \frac{\int_0^1 s^{X(q)-1} s\ ds}{\int_0^1
  s^{X(q)-1}\ ds} = \frac{X(q)}{1+X(q)} \;.
\end{equation} 
This relation is only meaningful as $q\to 0$; in this limit $f_p(q)$
has a non-trivial $q$ behavior and goes to $p$. When $q>q_\text{EA}$,
$X_J(q)=1$ for all samples, and $f_p(q)=1$.

The distribution of Eq.~(\ref{eq:pow}) is unusual in statistical
physics, since not only are the average, the median and the most probable
value different, but also the average and the median scale
differently as $q\to 0$.  Any ``nice'' distribution would have an
average and a median that would  scale in the same way, possibly
with different exponents; in this case $f_p(q)$ would go to one as
$q\to 0$.

\section{MONTE CARLO RESULTS AND THE MEAN-FIELD PREDICTIONS}
\label{sec:MC}

In this section we use numerical data obtained from parallel tempering
simulations for the SK and \EAI model to study the cumulative overlap
distributions and to test the mean-field predictions.

\subsection{The Monte Carlo simulations and the overlap databases}

Our work is based on a numerical database of low-temperature
thermalized configurations both for the three-dimensional \EAI
model~\cite{janus:10} and for the SK
model~\cite{billoire:00,billoire:02,aspelmeier:08}.  The three
dimensional configurations have been obtained on the Janus
computer~\cite{janus:09,janus:12b}. We have configurations and overlap
measurements for lattice sizes $L=8$, $12$, $16$, $24$ ($4000$
disorder samples) and $L=32$ ($1000$ disorder samples).  The lowest
temperature simulated at $L=32$ is $T=0.703$ (a recent
estimate of the critical temperature for this model is
$T_\text{c}\simeq1.1019(26)$~\cite{janus:13}).  Our SK database consists of $1024$
disorder samples of systems with total number of spins up to $N=4096$
and a lowest temperature $T=0.4$ (in this model $T_\text{c}=1$).

From this database we can obtain $N_q$ overlap measurements for each
disorder  sample, which we use to compute the $X_J(q)$.  The overlap
values either have been obtained directly during the numerical
simulation that has generated the spin configurations or have been
measured from stored thermalized configurations.  $N_q$ is of order
$10^6$ (but for the \EAI data with $N\leq 16$ where it is of order
$10^{5}$).

In general this amount of information is sufficient to obtain a
reasonable estimate for the single-sample overlap distributions.
However, at small values of $q$, where small values of $X_J$ occur
with high probability, we encounter a potentially severe statistical
problem.  The Monte Carlo method estimates $X_J$ as a population
(i.e., an integer number) divided by $N_q$. If the true value of $X_J$
is much smaller than $1/N_q$, the Monte Carlo estimate will be
exactly zero with a high probability (namely $\approx 1-X_J N_q$).

This effect is particularly important when estimating
$f_p(q)=I_p(q)^{X(q)}$, which is sensitive to small changes in the
values of the quantile $I_p(q)$.  Although small absolute
uncertainties in the determination of $X_J(q)$ do not have a large
impact on the estimate of $X(q)\equiv \overline{X_J(q)}$, measuring
either a very small or a null quantile can make a huge difference in
computing $f_p(q)$, which at small $X(q)$ can result in either a
quantity of order $1$ or exactly zero.  For example take 
$X(q)\sim 0.1$, roughly corresponding to $q\sim 0.2$, and try to
estimate $I_p(q)\equiv f_p(q)^{1/X(q)}\sim p^{1/0.1}$.  If $p=\frac12$
we are trying to estimate a number of order $10^{-3}$, that only
requires  limited statistics.  Things are different for
$p=\frac1{10}$, where we would be trying to estimate a number of order
$10^{-10}$, that would need of the order of $N_q=10^{10}$
measurements.  In general, for a given value of $p$ and $N_q$, we are
confident in the quality of the data down to $X(q)\sim
\log(p)/\log(1/N_q)$. For example, with the largest size we simulated,
the ``safe'' limits are roughly $X(q)\gtrsim 0.05$ for $p=\frac12$ and
$X(q)\gtrsim 0.15$ for $p=\frac1{10}$. Thus, although the
relation~(\ref{eq:fpq}) gives a more accurate prediction for lower
quantiles, for low values of $p$ the small-$q$ region cannot be
analyzed with acceptable accuracy.

We analyze the relevant physical quantities at the lowest temperature
available ($T=0.4$ for the SK model and $T=0.703$ for the \EAI model).
Due to the different critical parameters of both models, these two
temperatures are actually reasonably close in physical terms, as discussed in
Ref.~\onlinecite{billoire:13}.  In particular, the best extrapolations
down to $P(q=0)$ are very similar for both models at these temperatures,
implying that for small $q$ their $X(q)$ are very close.

\begin{figure}[t]
\begin{center}
\includegraphics[width=0.85\columnwidth]{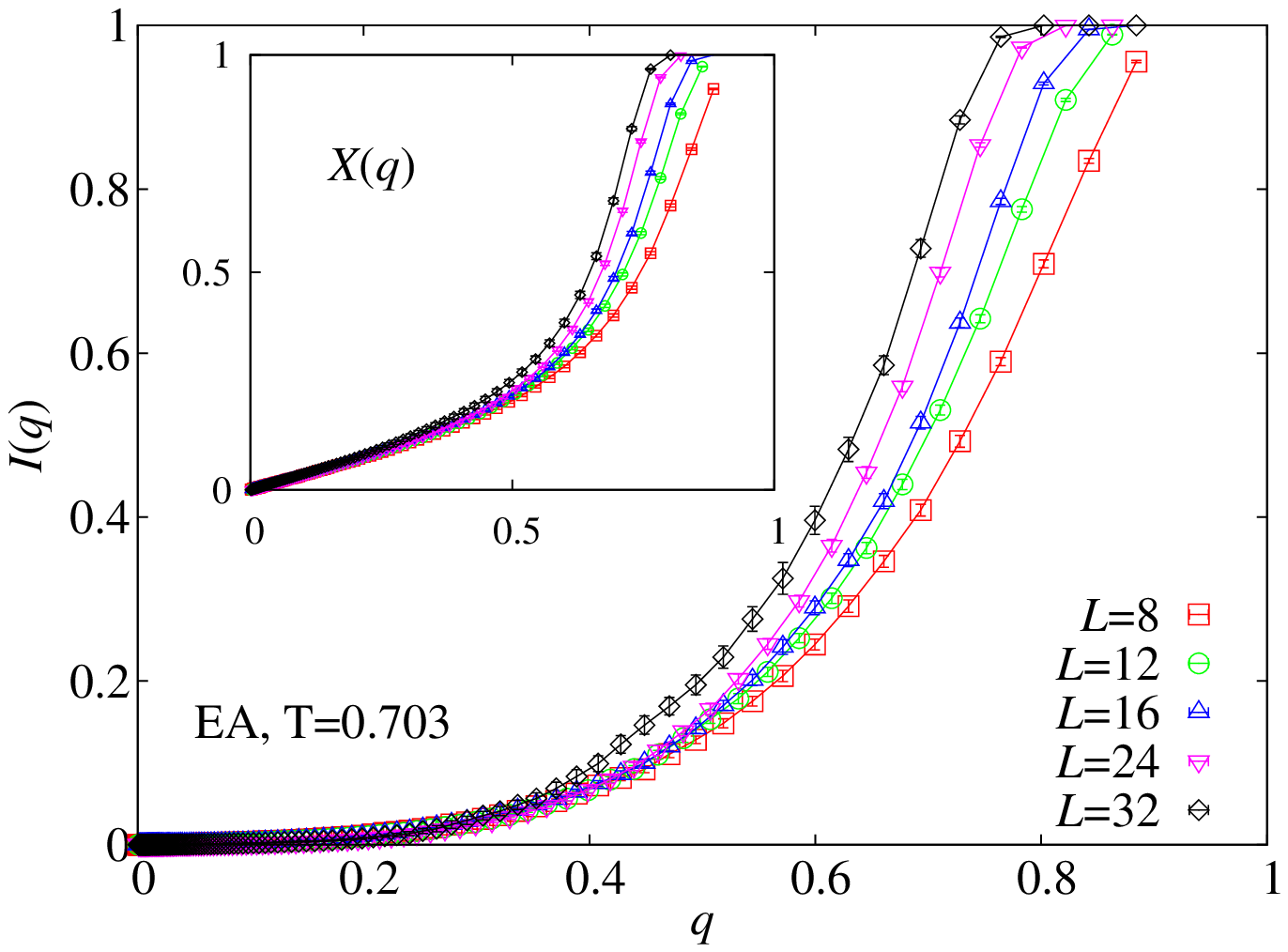}\\
\includegraphics[width=0.85\columnwidth]{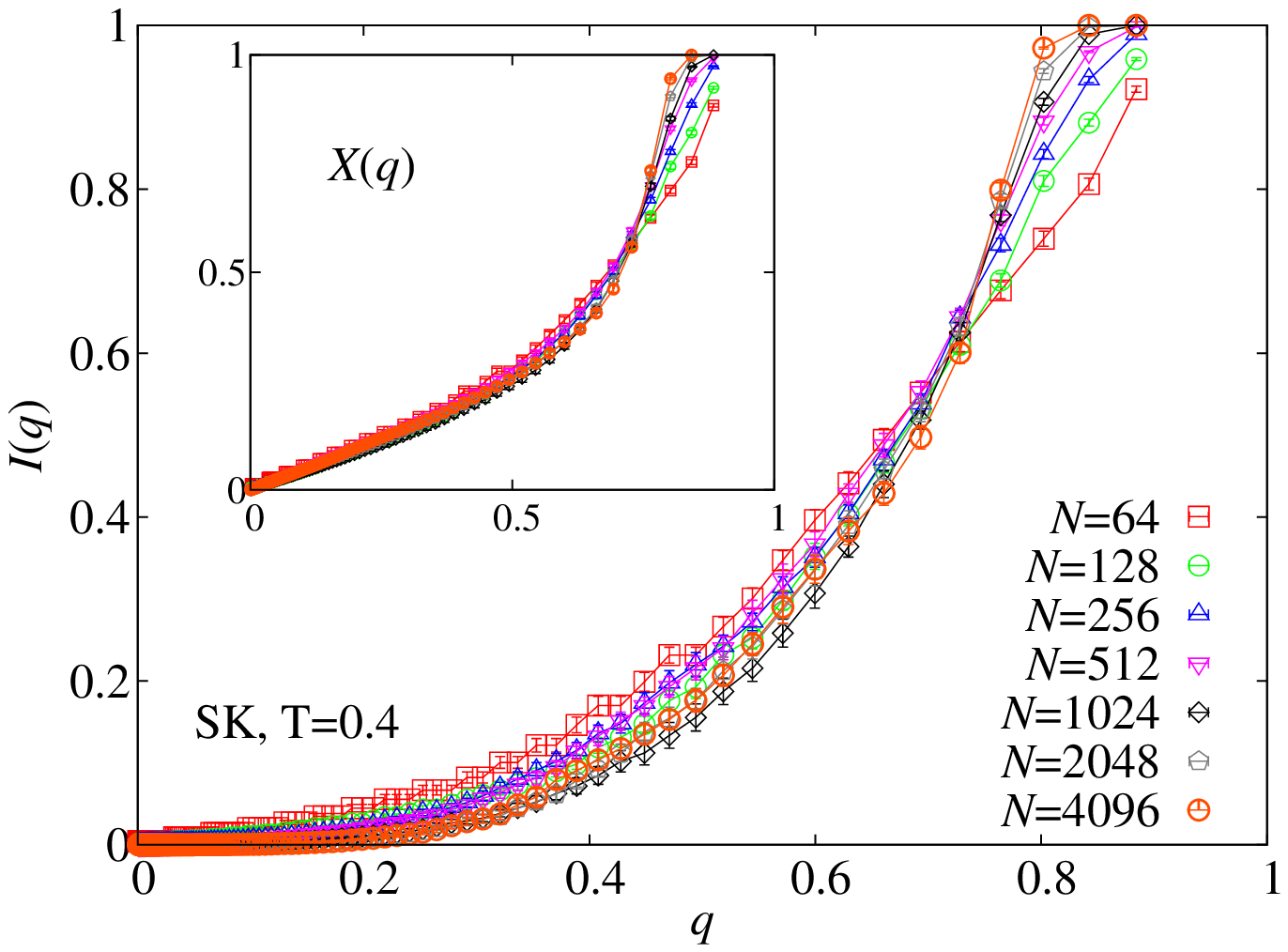}
\caption{(Color online) Top: the median $I(q)$ of the cumulative
  overlap distribution as a function of $q$ for the $3D$ \EAI model
  with temperature $T=0.703$. The inset shows the cumulative
  distribution $X(q)$ as a function of $q$. Bottom: the same curves
  for the SK model, with $T=0.4$.}
\label{fig:I_X_EA_SK}
\end{center}
\end{figure}

The \EAI model has not been simulated at exactly the same temperatures
for the different lattice sizes we have studied (as opposed to the SK
model where we have a consistent set of temperature values). Usually,
the temperature dependencies are smooth and we can safely interpolate
the data to the lowest temperature of the largest system
($T=0.703$). This procedure is followed in Fig.~\ref{fig:I_X_EA_SK}.
For more complicated (and noisy) quantities, such as $f_p$, given the
sensitivity to small errors, this procedure may be dangerous.
Therefore, in the following we have only analyzed the larger system
sizes, where the simulated temperatures where already very close to
the value we have for $L=32$, $T = 0.703$ ($T=0.697$ for $L=24$ and $T
= 0.698$ for $L = 16$) and the possible error due to the temperature
variation is negligible (see also  Refs.~\onlinecite{janus:11,janus:10}).
In any case, as we shall see, even if the $q$-behavior of our studied quantities
is $T$-dependent, we do not expect the $X$-behavior to be, so the precise
temperature will not be relevant in the rest of the paper (we just need to ensure
that we are as free as possible from critical effects, hence our choice of the 
lowest available temperatures).

If not specified otherwise, the error bars in the plots have been obtained
with a  bootstrap analysis~\cite{efron:94}.

\subsection{Numerical results for the EAI and SK models}

Let us start by considering $I(q)$ and $X(q)$ as a function of $q$,
which we show for the \EAI and the SK models in
Fig.~\ref{fig:I_X_EA_SK}. In the SK model both $I(q)$ and $X(q)$
converge nicely to some limiting curve when $N$ increases.  The major
source of finite-size effects is apparently the well-known shift of
$q_\text{EA}$ for increasing $N$ values~\cite{janus:10,janus:10b,aspelmeier:08}. The
average $X(q)$ is linear in $q$ at the origin as expected. For small
values of $q$ $I(q)$ is much smaller than $X(q)$, as expected
from~(\ref{eq:powCum}). The comparison with the plots of
Ref.~\onlinecite{middleton:13} for droplet-like models shows a marked
difference, since there $I(q)$ is identically zero for small values of
$q$. The study of the median of the distribution of $X_J(q)$ as a
function of $q$ does distinguish clearly between droplet and RSB
mean-field behavior. Trading the average for the median does make the
analysis more clear cut.
 
The \EAI data are very similar to those obtained from the SK model:
both $I(q)$ and $X(q)$ for increasing values of $N$ nicely converge to
some limiting curve (in agreement with the fact~\cite{janus:10} that
$P(0)$ depends only very weakly on $L$). The limiting curve for $X(q)$
is linear at the origin. $I(q)$ is much smaller than $X(q)$ but it is
definitely not identically equal to zero at low $q$, unlike in the
droplet-like models of Ref.~\onlinecite{middleton:13}. In conclusion
the study of the $q$ behavior of the median of the $X_J(q)$
distribution gives strong support to a RSB scenario for the 3D \EAI
model.

\begin{figure}[t]
\includegraphics[width=0.85\columnwidth]{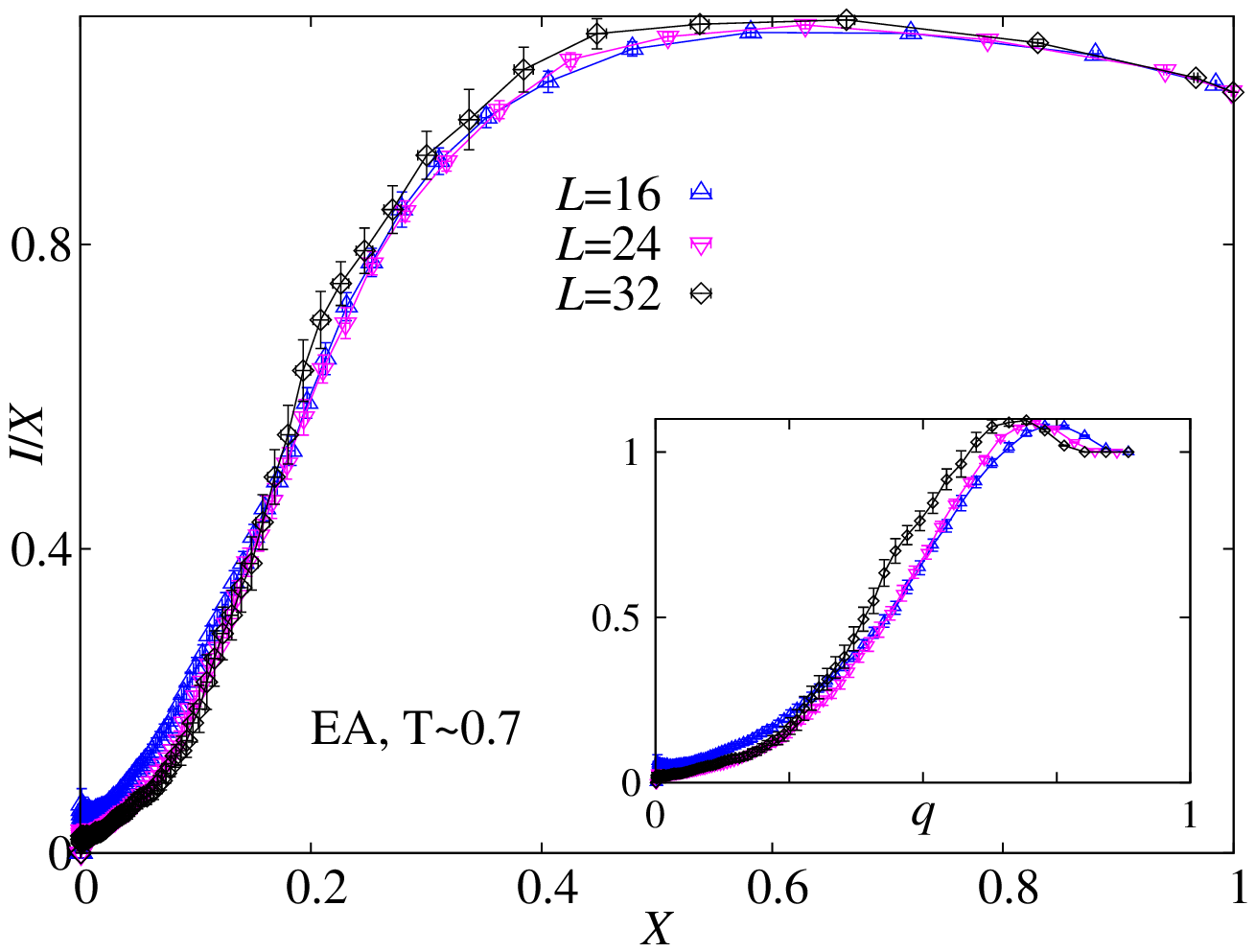}\\
\includegraphics[width=0.85\columnwidth]{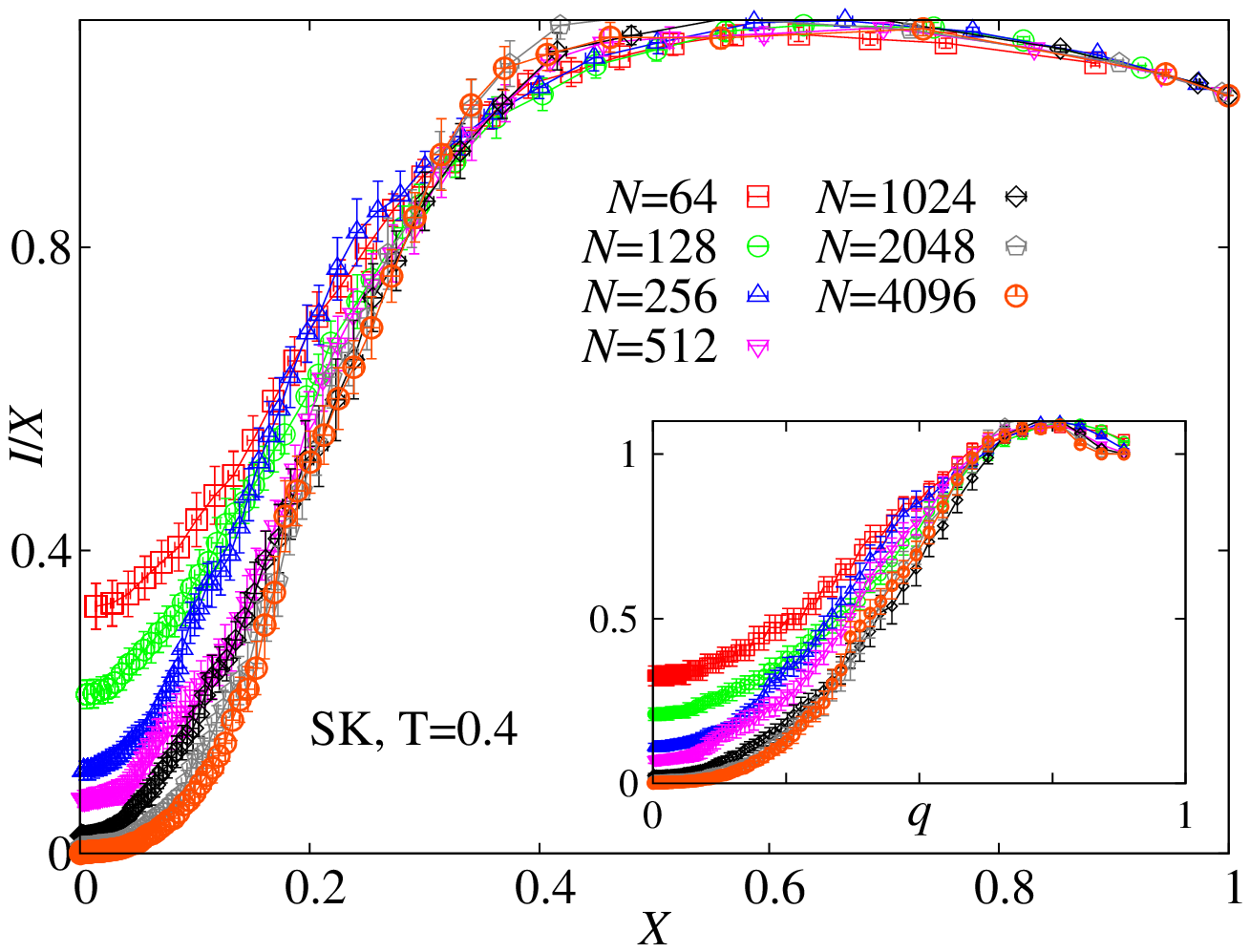}
\caption{(Color online) The median  to mean ratio $I(q)/X(q)$ as a
  function of $X(q)$ (main plots) and as a function of $q$
  (insets).  Top: 3D \EAI data with $T\sim 0.7$; bottom: SK data with
  $T=0.4$. Note the much lager finite-size effects in the SK case.}
\label{fig:R_EA_SK}
\end{figure}

In Fig.~\ref{fig:R_EA_SK} we show the ratio $I/X$ as a function of $X$
(main plots) and of $q$ itself (insets) for both the \EAI and SK
models.  The SK data show very strong finite-size effects as $q$ and
$X$ go to zero. In contrast the \EAI data show little finite-size
effects. While this difference remains to be understood, the upshot is
that in both the \EAI and SK cases, the ratio $I/X$ is vanishing for
small $X$ and for large system sizes, as expected from the RSB
picture.

\begin{figure}[htp]
\includegraphics[width=0.85\columnwidth]{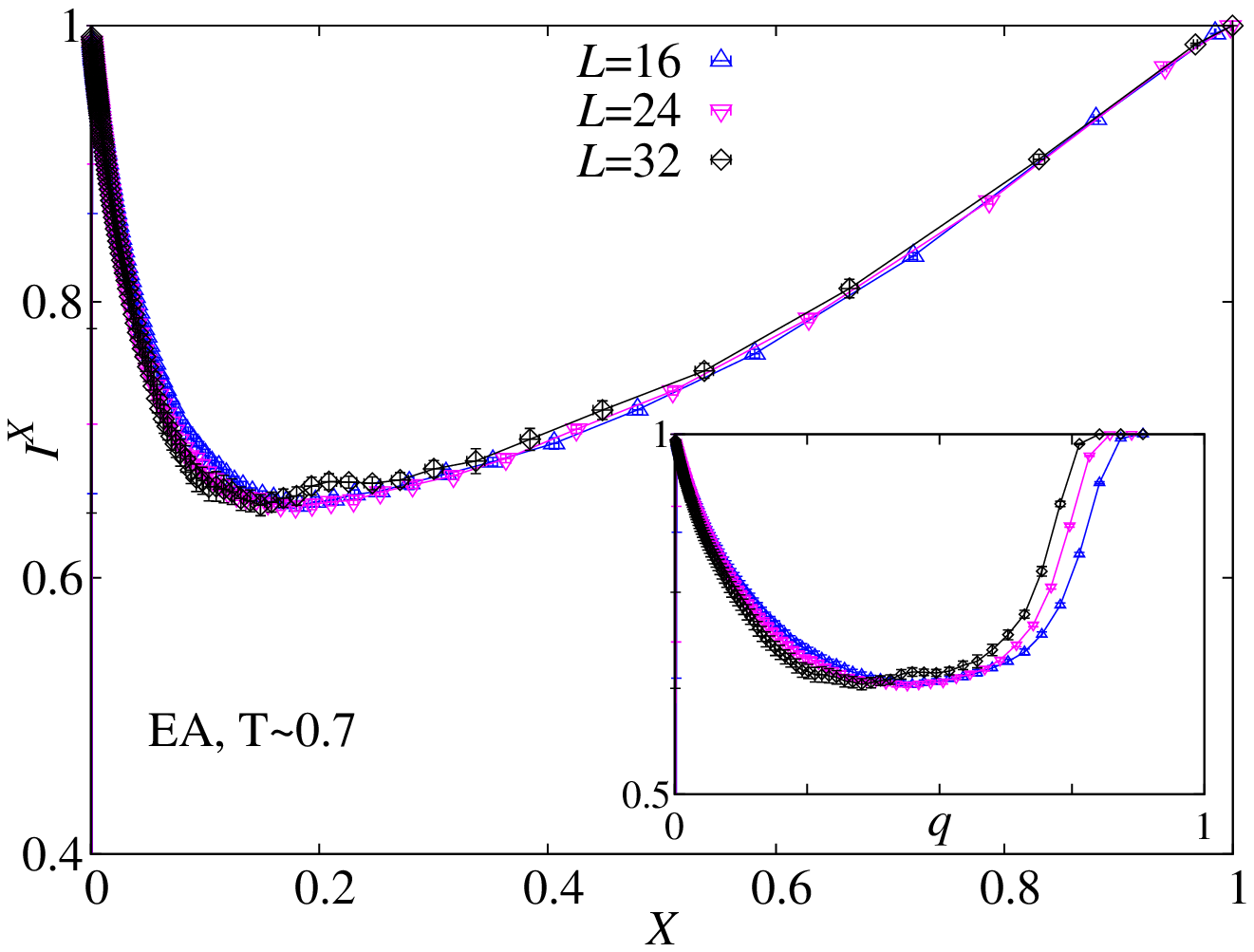}\\
\includegraphics[width=0.85\columnwidth]{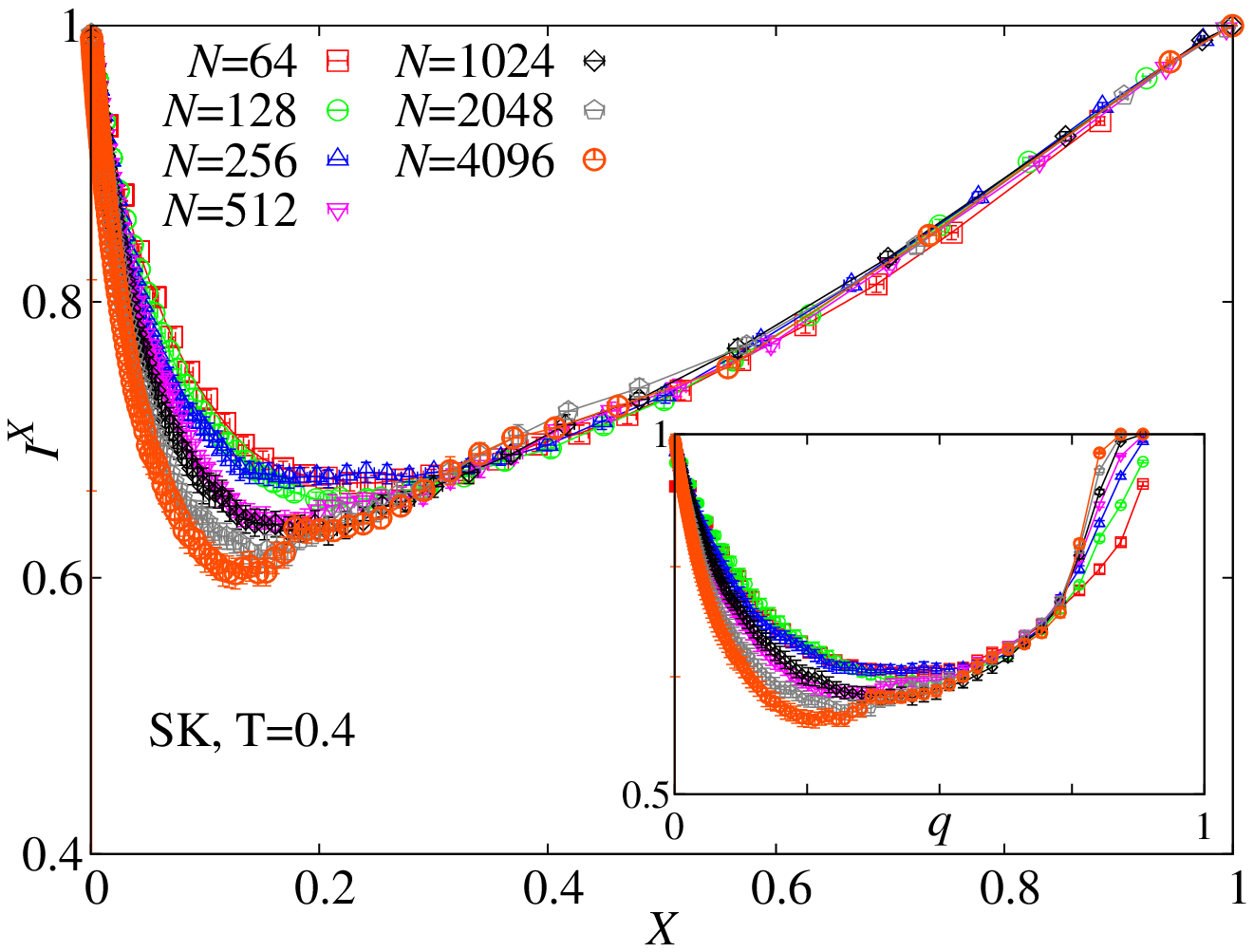}
\caption{(Color online) $f(q)\equiv I(q)^{X(q)}$ as a function of
  $X(q)$ (main plots) and as a function of $q$ (insets).  Top: $3D$
  \EAI model with $T\sim0.7$. Bottom: SK model with $T=0.4$. Note the
  much larger finite-size effects in the SK case.}
\label{fig:f_EA_SK}
\end{figure}
\begin{figure}
\includegraphics[width=0.85\columnwidth]{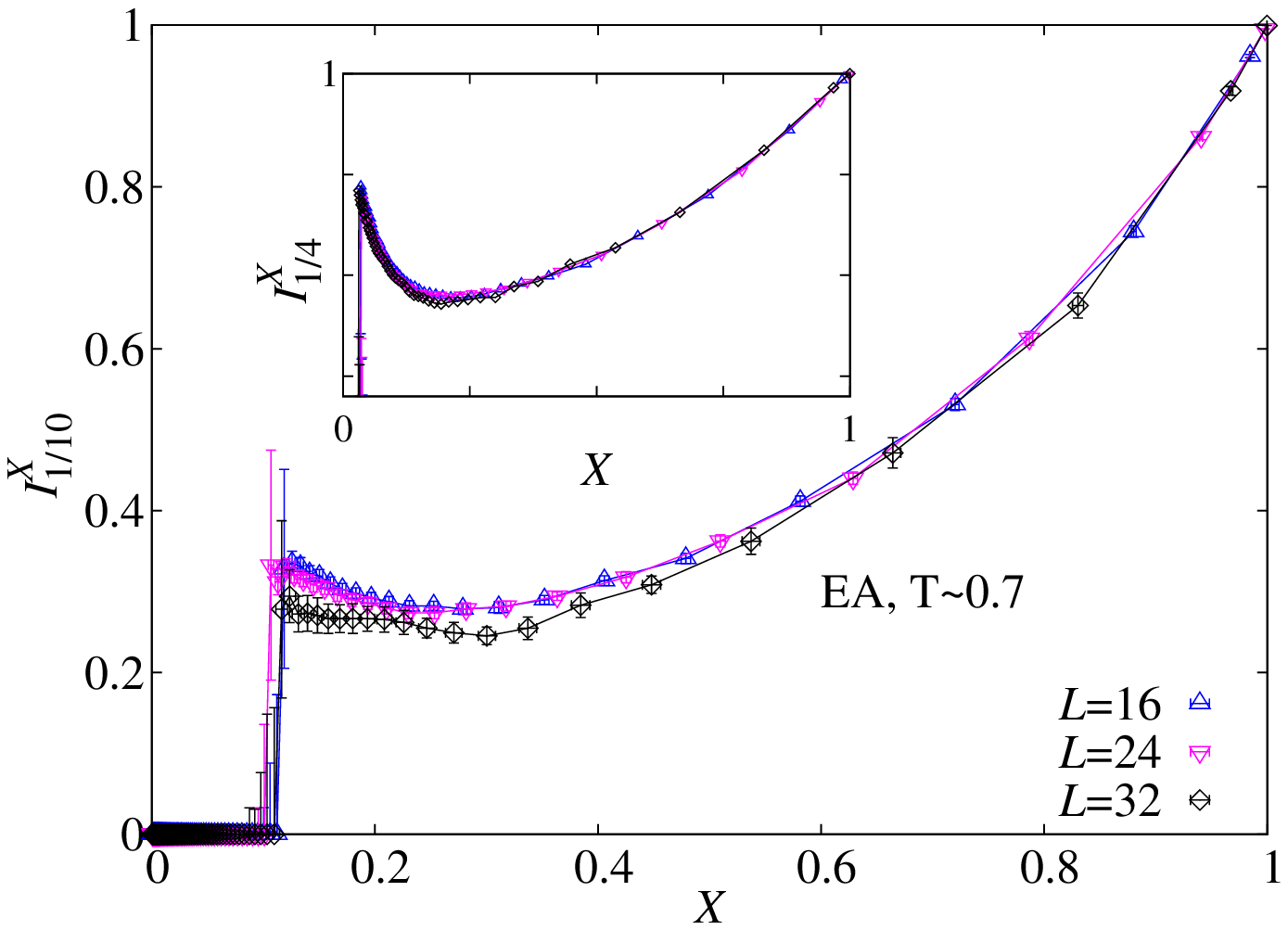}\\
\includegraphics[width=0.85\columnwidth]{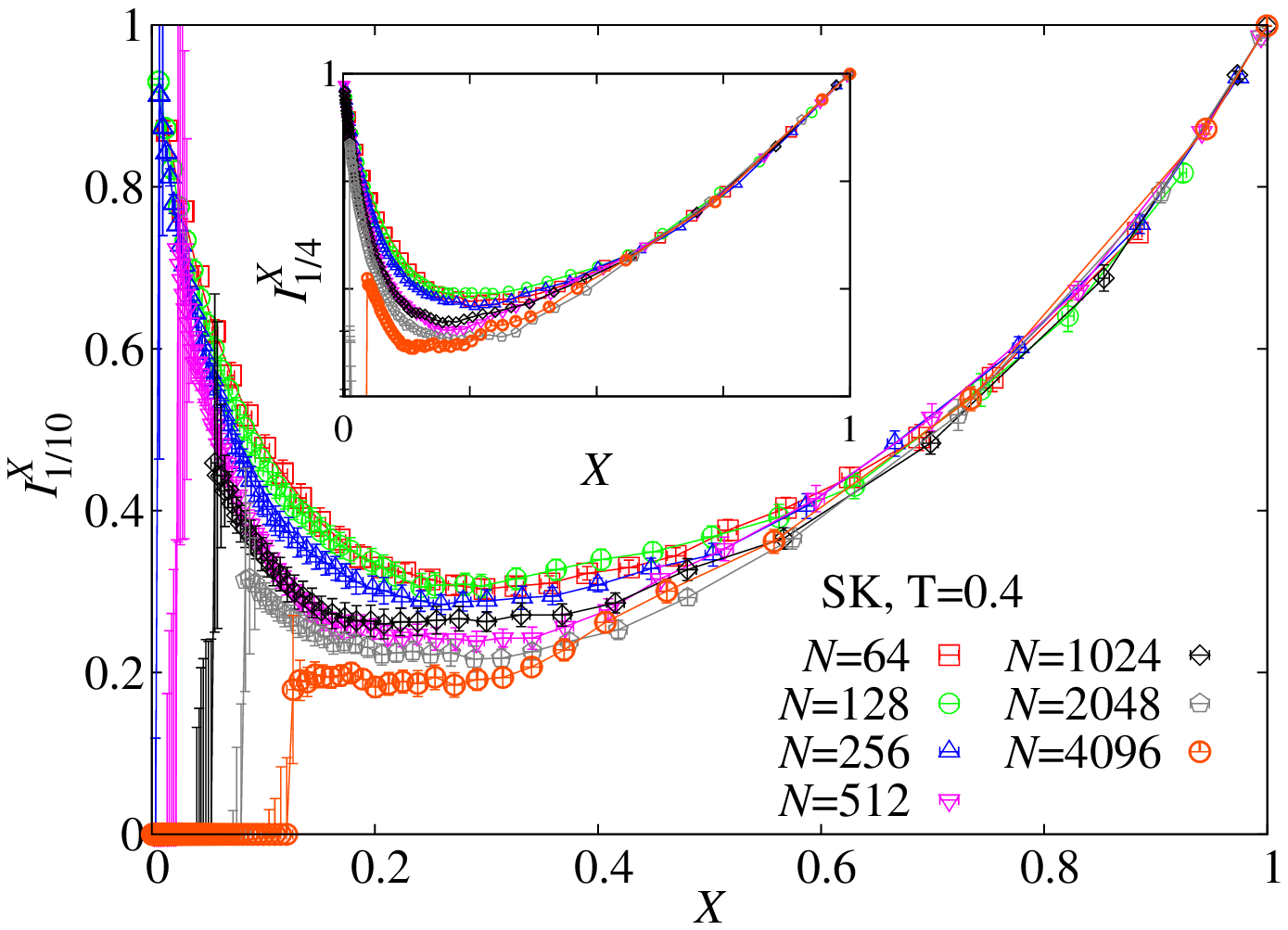}
\caption{(Color online) The quantities $f_{1/10}$ (main plots) and
  $f_{1/4}$ (insets) as a function of $X$.  Top: $3D$  \EAI data with
  $T\sim0.7$. Bottom: SK data with $T=0.4$.}
\label{fig:fp_EA_SK}
\end{figure}

In Fig.~\ref{fig:f_EA_SK} we show $f(q)\equiv I(q)^{X(q)}$ as a
function of $X(q)$ (main plots) and as a function of $q$ (insets) for
both the \EAI and the SK models.  The interpretation of the SK data is
clear: the data for increasing system sizes converge toward a smooth
limiting curve, whose $q\to 0$ (or $X\to 0$) limit is compatible with
the expected value $\frac12$. The convergence fails when finite-size
effects become important: for any given value of $N$ there is a
crossover value $q^*=q$ below which $f$ enters a finite-size regime
and goes to one as $q$ and $X(q)\to 0$. This  $q^*$ goes to zero as $N\to\infty$.
In other words, for each value of $N$ there is a value of $q$ below
which the finite-size broadening of the peaks in the overlap
distribution cannot be neglected, and the distribution of the $X_J$'s
becomes a ``nice distribution'' whose median and average scale in the
same way (possibly with different exponents) when $q\to 0$.  We can summarize the situation by saying that
$\lim_{q\to 0}\lim_{N\to_\infty}f=\frac12$, but that at fixed, finite
$N$ $\lim_{q\to 0} f=1$.

The overall situation is very similar for the \EAI model (top part of
Fig.~\ref{fig:f_EA_SK}). Here again we have a function decreasing, for
decreasing $X$, down to a small value of $X$ and eventually increasing to one.
Here the emergence of the thermodynamic behavior looks slower than for the SK
model; this makes it difficult to estimate the infinite-volume limit of $f$ for
$X\to 0$, but qualitatively it is crucial that we have the same kind of
behavior than in the SK model, in the same range of $p$ values.

In Fig.~\ref{fig:fp_EA_SK} we show $f_p(X)$ for the lower-order
quantiles $p=0.25$ and $p=0.1$ for both the \EAI and SK models, that
turn again, in both cases, to be consistent with RSB predictions.  The
data can be interpreted exactly like the $p=1/2$ data, with the
difference that now the low $q$ data are severely affected by the
finite $N_q$ bias presented above. All points where $I_d\ll 1/N_q$ are
cut off, and $f_p=0$ in the whole region at the left of some size and
quantile dependent threshold.  As a collateral damage due to a
statistical bias (and not to a physical effect) the finite-size rise
of $f_p$ towards $1$ as $q\to0$ (or $X\to 0$) is lost. Again in the
\EAI model we observe a far weaker volume dependence than in the SK
model, and we only detect slow and weak hints of the emergence of the
thermodynamical behavior of the system.

\subsection{Comparison of the numerical results with the mean-field expectations}

\begin{figure}[htp]
\includegraphics[width=0.95\columnwidth]{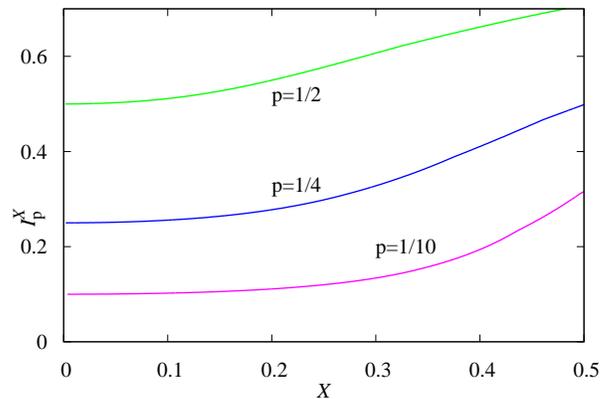}\\
\caption{(Color online) Prediction for $f_p=I_p(q)^{X(q)}$ as a
  function of $X(q)$ using the self-consistent approximation of
  Eq.~\ref{eq:powReg}. Curves are drawn for $p=\frac12$ (i.e., for the
  median), for $p=\frac14$ and for $p=\frac1{10}$.}
\label{fig:fp_reg}
\end{figure}
\begin{figure}[htp]
\includegraphics[width=0.95\columnwidth]{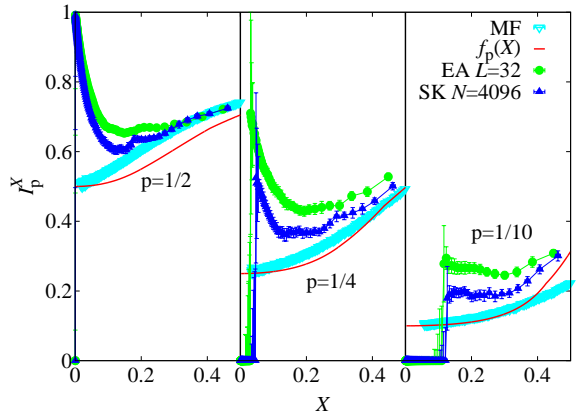}\\
\caption{(Color online) We compare the self-consistent approximation
  of Eq.~\ref{eq:powReg}, $f_p(X)$, to the data obtained for the largest lattice
  sizes both for the \EAI and the SK models. We also include the mean-field
  prediction, as constructed from a weight-generation method (see text and~\cite{marinari:98d}).
  We show on the left
  $p=\frac12$ (i.e., the median), in the center $p=\frac14$ and on the
  right $p=\frac1{10}$.}
\label{fig:fp_reg2}
\end{figure}

In this subsection, we compute the function $f_p(q)$ in the mean-field
theory beyond the simplest approximation used in the previous subsection (namely $f_p(X)=p$). We consider two different approximate methods, 
and  compare the results to  our numerical data.
As we have discussed before, the prediction of
Eq.~(\ref{eq:pow}) only holds as $q\to0$ (and it is not even self-consistent).
As a modest improvement we can make it self-consistent while keeping the
correct $q\to0$ behavior, writing $\mathbb{P}(s) = A(X)s^{X-1}+B(X)$, where
$A(X)$ and $B(X)$ are fixed by the normalization of $\mathbb{P}(s)$ and by
self-consistency (the relation analogous to Eq. (\ref{eq:self})). For instance, 
when $B(X)=0$  then $A(X)=X$ and $f_p=p$. This can be
done at least when $X<0.5$, with the result:
 \begin{equation} \label{eq:powReg}
\mathbb{P}_C(s) = \frac{s}{1-X} \left\{ s^{X-1} \left( 1-X-2X^2 \right) +2 X^2
\right\}\;.  \end{equation} 
Now computing $\mathbb{P}_c(s)$ at $s=I_p$ gives, for
$f_p\equiv I_p^X$, the equation
 \begin{equation} \label{eq:fp2}
\frac1{1-X}\left(\left(1-X-2X^2\right)f_p+2 X^2 f_p^{\frac1X}\right) = p\;.
\end{equation} 
We show in Fig.~\ref{fig:fp_reg} the functions $f_p(X)$ as
obtained by this simple modification of Eq.~(\ref{eq:powCum}): one has, as
expected, $\lim_{X\rightarrow0}f_p=p$, and $f_p$ is a monotonic function of $X$
which is almost flat near $X=0$.

There is an alternative approach to the estimation of $f_p$. Indeed, the
function $\mathbb{P}(s=X_J)$ is actually known in mean field for a given value
of $X$ (see~\cite{mezard:87, mezard:85b}). The full equations are complicated,
but there is a simple numerical method~\cite{marinari:98d} that can be used to
sketch the behavior of $I_p(X)$. Essentially, we take advantage of the
ultrametric structure of the spin-glass phase to group the states in clusters
at any value of $X$.  We consider a system where $M$ such clusters are allowed,
each with a weight $w_a = C \exp(-f_a)$, where $C$ is such that $\sum_a w_a =
1$. The $f_a$ are i.i.d random variables distributed according to $p(f_a) = B
\exp(X f_a)$. We can then use this set of weights to compute the $X_J$ for a
given sample. This method provides no relation between the $X_J$ for  a fixed
sample at different levels of $X$, so it cannot be used to generate the full
$P_J(q)$, but it is useful to sketch the behavior of $I_p(X)$ and, hence, of
$f_p$ (at least for not too small values of $X$: for $X\ll1$ the method
involves the sampling of huge or tiny numbers and the numerical computation breaks
down) \footnote{We could also have extracted the $w_a$ from the known exact
distribution emerging from the stick-breaking process reported
in~\cite{derrida:87}. However, this construction is already almost
indistinguishable from the numerical method summarized above for $M=16$ and
suffers from the same small-$X$ problems.}.

In Fig.~\ref{fig:fp_reg2} we show  the predictions of the self-consistent
approximation together with the weight-generation method and with
the numerical data for the largest lattice
size for both our models. The qualitative agreement is very reasonable, 
specially for the small-$X$, small-$p$ sector.

\section{THE DROPLET PICTURE AND THE TOY MODELS}
\label{sec:TOYS}
Thus far we have described the expected scaling behavior of the
cumulative overlap distribution in the RSB picture, and we have
checked that our numerical data are compatible with it, both for the
SK and for the \EAI model.  The next step would be to make the
analogous test comparing to the expected scaling behavior in the
droplet theory.  However, in this picture, a specific prediction for
the finite-size behavior of these quantities is not available.
Therefore, in order to test the hypothesis of a droplet-like behavior
of the \EAI model, we introduce several droplet-like
(single-state) toy models, and we will compare their behavior with the
one emerging from our numerical data. In addition, we  introduce
several many-state toy models (representing a simplified mean-field
picture), in order to discuss to which extent the validity of
\eqref{eq:pow} is an unavoidable consequence of the existence of a
non-trivial overlap distribution.

\subsection{Definitions of toy models}

\begin{itemize}
\item \textbf{Models \boldmath $D2$ and $D3$.} First, we
  consider a version of the toy model of Ref.~\onlinecite{hatano:02},
  described and studied in Ref.~\onlinecite{middleton:13}. For a
  system of size $L$, one defines a \textit{sample} as a set of
 independent active droplets. These are group of spins that always keep their relative orientations but may flip as a
  whole, with probability $\frac12$.  These droplets are, in this
  model, quenched in size, and their distribution embeds the quenched
  disorder that characterizes the model: each droplet has a fixed,
  defined size, that does not change in time. On such a droplet sample
  one studies thermal averages where droplet signs change, as we said,
  with probability one half, allowing us to compute, among others, the
  overlaps in a given sample.    The number $n_v$
  of active droplets of size  $v$  is Poisson
  distributed, with mean  $n_v  =c\,v^{-x}L^D$, where $x=2+\theta/D$, so
  that the average number of droplets of size $\ell^D$ scales as
  $\ell^{-D-\theta}$, as expected in the droplet scenario. In this toy model
  droplets are not defined relative to a lattice, and the dimension $D$ is just a parameter.
  We proceed  in two phases. We first fix a sample by defining
  the droplets, and second we dynamically change their sign, computing
  in this way expectation values for a given sample. We generate a
  sample by extracting numbers of droplets with size up to $L^D/2$
  from the Poisson distribution (for very small $L$ values it can
  happen that $\sum_{v=1}^{L^D/2} v n_v > L^D$: in this case we
  discard the sample and try again) and we add to it an extra (large)
  droplet of size $L^D - \sum_v v n_v$.  Following
  Ref.~\onlinecite{middleton:13}, we consider two versions of the
  model: the model \emph{D}$2$, where $D=2$, $\theta=0.5$ and $c=0.1$
  (mimicking a two-dimensional droplet system), and the model
  \emph{D}$3$, where $D=3$, $\theta=0.21$ and $c=0.0375$ (for a
  three-dimensional version of the model).  With this choice of
  parameters in the $D=2$ model the ``large'' droplet occupies in
  average close to $54\%$ of the lattice, while in the
  $D=3$ case it takes close to $44\%$ of the lattice.

\item \textbf{Model \boldmath $PK0$.} Here we define the model by
  assigning the overlap probability distribution $P_J(q)$.  We take for
  $P_J(q)$ a pair of Gaussian distributions with fixed width $\sigma$ centered at random positions
  $\pm q_J$:
  \begin{equation} P_J(q) = \frac{1}{\sqrt{8\pi \sigma^2}}
    \left\{e^{-\frac{(q+q_J)^2}{2\sigma^2}}
      +e^{-\frac{(q-q_J)^2}{2\sigma^2}}
    \right\}\;,
    \label{eq:pjqPK0}
  \end{equation}
  By varying $\sigma$ we can mimic the
  broadening of the distribution due to finite-size effects.  The
  value of the peak locations is extracted from a hard-tail
  probability density:
  \begin{equation}
\begin{split}
    \mathcal{F}(q_J;\epsilon) & \propto 
    \epsilon^{-1}\exp{\left(-\frac{1}{1-\left((q_J-q_{J_0})/
    \epsilon\right)^2}\right)}\;,\\
    \mathcal{F}(q_J;\epsilon) & =  0\ \mbox{when}
    \ \left|q_J-q_{J_0}\right|\geq \epsilon
    \label{eq:fqJ}
\end{split}
  \end{equation}
  which behaves like a delta function $\delta (q_J-q_{J_0})$ in the
  $\epsilon \rightarrow 0$ limit.  In this over-simplified
  description, the two-state picture corresponds to the 
  $\epsilon \to  0$ limit of the model. In the following we will 
  take $q_{J_0}=0.7$.

\item \textbf{Model \boldmath $PK\lambda$.} In a slightly more
  elaborated version of \textit{PK}0, we allow for more peaks, besides
  the one at $q_{J_0}$, to contribute to $P_J(q)$ for $q_{J_0}\geq q\geq
  0$. Here a sample has a random number $N_J$ of secondary peaks at
  locations $q_k$, with $k=0,\dots,N_J-1$.  The number of secondary
  peaks $N_J$ is a Poisson-distributed random variable with mean
  $\lambda$. The secondary peaks locations are uniformly distributed
  in the interval $[0,q_{J_0}]$, and the weights of the primary peak
  $W_{N_J}$ and of each of the $N_J$ secondary peaks $W_k$ are
  i.i.d. with  uniform probability density.  $P_J(q)$ is then
  a sum of pairs of Gaussian distributions:
\begin{equation}
\begin{split}
P_J(q)  = 
\sum_{k=0}^{N_J}\frac{W_k}{\sqrt{8\pi \sigma^2}}&
\left\{e^{-(q+q_k)^2/2\sigma^2}\right.\\ &+
\left.e^{-(q-q_k)^2/2\sigma^2}\right\}\;.
\end{split}
\label{eq:pjqPKl}
\end{equation}
where $q_{N_J}=q_{J_0}$. Also here we take $q_{J_0}=0.7$.
This model can be seen as a \textit{many-states} version of the \emph{PK}$0$
model proposed above.  For a given disorder realization
the quenched disorder is given by the positions of the peaks and their
weights, i.e., 
\begin{equation}
\nonumber
\left\{J\right\}
=
\left\{
q_J,W_{N_J},q_{k=0,\dots,N_{J}-1},W_{k=0,\dots,N_{J}-1}
\right\}\;.
\end{equation}
$X_J(q)$ 
can be easily computed as a sum of error functions.

\item \textbf{Model \boldmath $UB\lambda$.} Our last toy model uses a
  random branching process~\cite{parisi:93} to construct hierarchical
  trees of states. Starting from the root node at $X=0$, and
  incrementing $X$ in small steps $\delta X$ up to $X_M$, we allow any
  branch to bifurcate randomly at each step with a given fixed
  probability $\frac{\lambda\;\delta X}{X_M}$, such that any path from
  the root to any leaf has an average number of bifurcations equal to
  $\lambda$.  At a bifurcation, the weight of each new branch is
  assigned extracting two \emph{free energy} values $F_1$ and $F_2$
  and constraining the two weights $w_i \propto \exp(-F_i)$ to sum up
  to the weight of the ancestor, as described in
  Refs.~\onlinecite{mezard:85b,parisi:93}.  In order to map $X$ values
  to $q$ values, we simply take $X(q)$ to be a linear function of
  $q\in[0,q_J]$ with $q_J$ extracted as in the \emph{PK}$\lambda$ toy
  model. At each level of $X$, $X_J(q)=1-\sum w_i^2$, where the sum
  extends to all branches that have already spawned.  As $X_J(q)$ is a
  piece-wise constant function, the overlap probability density of the
  single sample is a sum of delta functions, that we smooth by
  Gaussian convolutions exactly as in the \emph{PK}$\lambda$ toy
  model.  It would take an infinite branching process and a precise
  knowledge of the function $X(q)$ to accurately reproduce the results
  from the mean-field theory~\cite{parisi:14}. Still this toy model
  has, by construction, interesting properties such as ultrametricity,
  non-self-averageness and a non-trivial average $P(q)$.  In our
  computation we set $X_M=X(q_J)=0.5$ and $q_{J_0}=0.7$. For
  $\lambda=0$ the two toy models \emph{UB}$0$ and \emph{PK}$0$
  coincide.
\end{itemize}
\begin{figure}[t]
\begin{center}
\includegraphics[width=0.85\columnwidth]{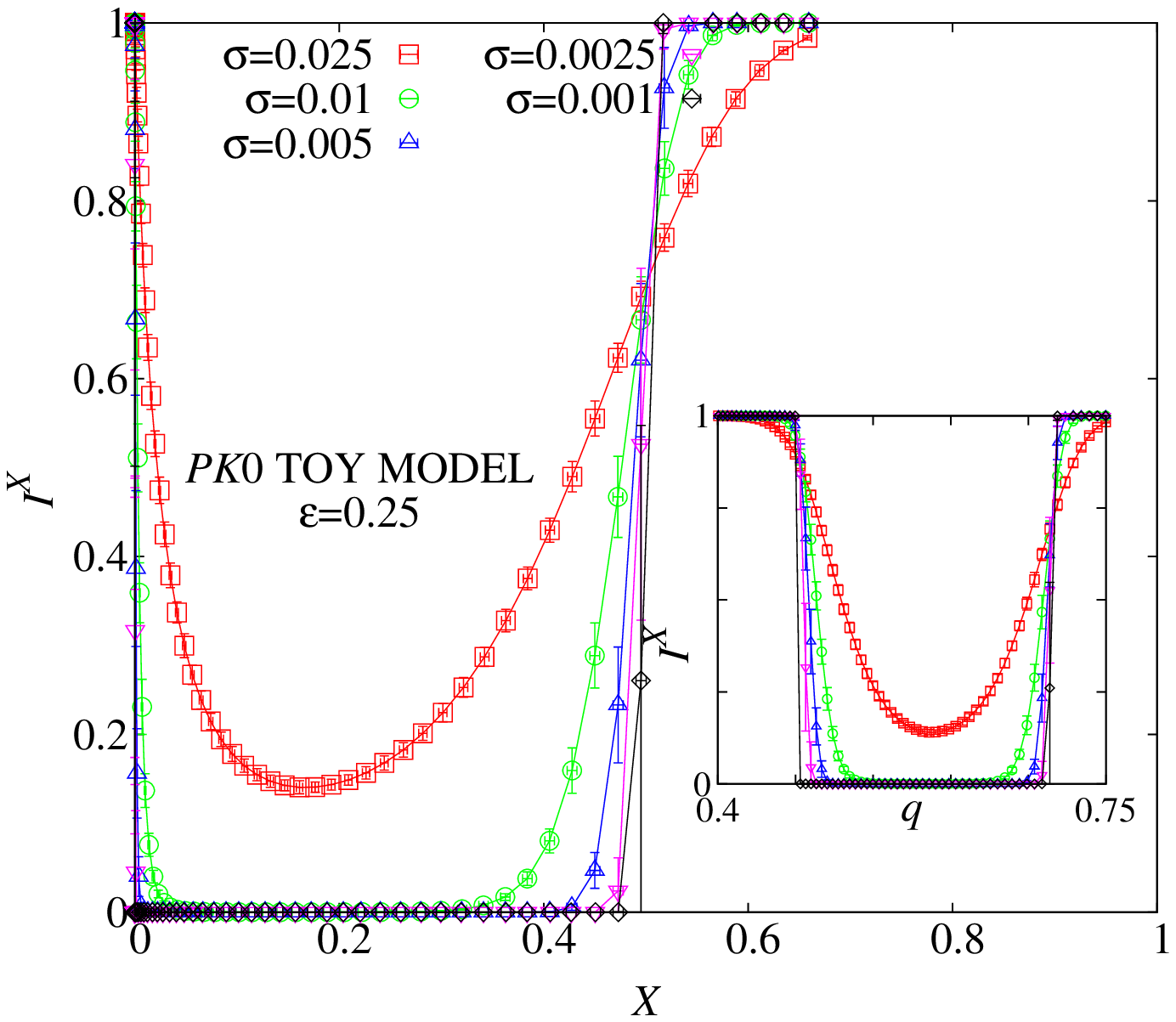}\\
\includegraphics[width=0.85\columnwidth]{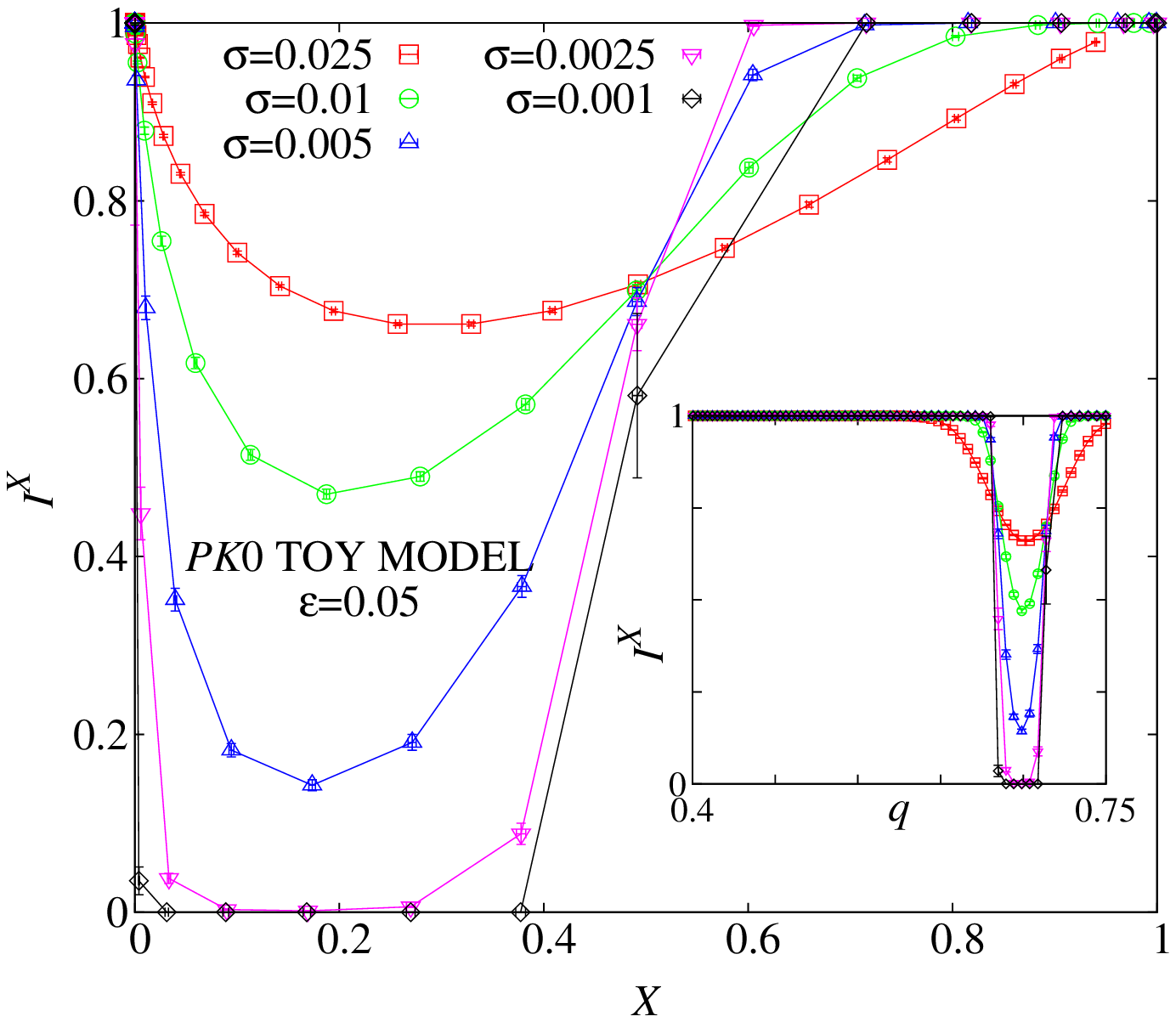}
\caption{(Color online) \emph{PK}$0$ toy model: $f\equiv I(q)^X(q)$ as
  a function of $X(q)$ (main plots) and as a function of $q$
  (insets).  $q_{J_0}=0.7$.
  Top: the width of the distribution of the position of the
  peak is $\epsilon=0.25$. Bottom: $\epsilon=0.05$.}
\label{fig:pk0}
\end{center}
\end{figure}
\begin{figure}[t]
\begin{center}
\includegraphics[width=0.85\columnwidth]{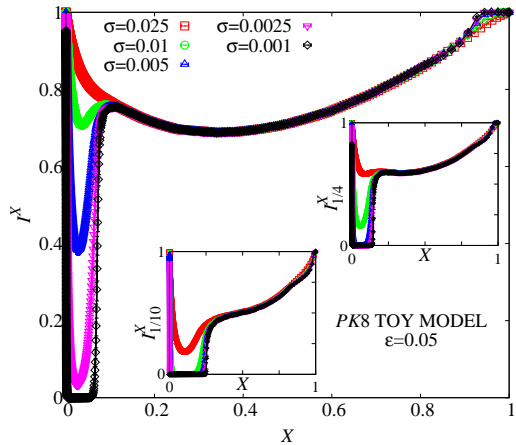}
\caption{(Color online) \emph{PK}$8$ toy model: $f_p$ as a function of
  $X$, for $p=1/2$ (main plot), $p={1/4}$ and $p={1/10}$ (insets).}
\label{fig:pk8}
\end{center}
\end{figure}

\subsection{Numerical results for the toy models}

Using these toy models we can check whether the results obtained in
Sec.~\ref{sec:MC} are really a consequence of the existence of a
RSB-like spin-glass phase or just a numerical artifact.

The \emph{PK}$0$ model reproduces, by tuning the $\epsilon$ and
$\sigma$ parameters, the trivial distribution one expects for a large
system in a droplet-like picture. When the distribution of the
self-averaging peak's position $q_J$ is a narrow delta 
($\epsilon\rightarrow 0$) around $q_{J_0}$, the $X_J(q)$ of most samples are all
very close, since their values depend almost exclusively on
$\sigma$. At intermediate overlap values $q\sim q_{J_0}-\epsilon$, the
few samples for which $q_J\lesssim q$ dominate the mean value, while
the median (or any other smaller quantile) stays small, and $f_p$ is
depressed. Outside such region, $I_p$ and $X$ are either both very
small (low $q$) or both of order one ($q\gtrsim q_{J_0}$), that
implies $f_p\sim 1$. As one can see from the insets in
Fig.~\ref{fig:pk0}, the region in which $f_p$, when seen as a function
of $q$, is significantly
different from unity shrinks when $\epsilon$ decreases. Since, by
construction, we cannot have samples with $q_J<q_{J_0}-\epsilon$, for
any (not too large) $\epsilon$ values all $X_J(q)$ are vanishing at
small but non-zero $q$ values when $\sigma\to 0$. As a function of
$X$, $f_p$ is almost one above $X(q_{J_0}+\epsilon)$, and almost zero
below $X(q_{J_0}-\epsilon)$.  As a function of $q$, we have a dip that
gets sharper and deeper as $\sigma$ decreases. The dip width shrinks
as $\epsilon$ gets smaller, and the values of $X$ cluster in two
narrowing regions around $X=0$
and $X=1$ respectively.  In \emph{PK}$0$ then, at large $\sigma$, $f$ is zero at
small $X$ but it is one at $X\sim 0$, or, in terms of the overlap, $f$
is zero in an interval of size $\sim\epsilon$ around $q_{J_0}$ and it
is one everywhere else.

The \emph{PK}$\lambda$ toy model adds a non-self-averaging
contribution to the overlap distribution. The secondary peaks can be
centered at any values of $q\lesssim q_J$, down to $q\sim 0$. When the
distribution of the self-averaging peak position $q_J$ gets narrower
(i.e., when $\epsilon$ gets smaller), a strong depression in the median
(and in the lower quantiles) is still possible at low $q$ and at small
$\sigma$, because the median of the position of the leftmost peak (the
median of the smallest $q_k$ value) has a finite distance from
$q=0$. As the number of allowed peaks grows, the dip does eventually
shrink to $q\sim 0$, but for large values of $\lambda$ the model
becomes trivial since it loses non-self-averageness. We show in
Fig.~\ref{fig:pk8} an example of what happens in the \emph{PK}$8$
model (where $\lambda=8$, i.e., there are in average eight secondary
peaks). The dip at low $X$ values is due to the fact that at low $q$
values there are no peaks.  When the number of peaks increases one
gets more peaks close to to $q=0$ and gets additional contributions to
$f_p$, which can become different from zero down to very low $X$
values. Apart from the dip, which is built-in in the toy model but is
related to finite-statistics artifacts in the spin-glass models (and
disappears for these models in the limit of an infinite number of
samples), \emph{PK}$8$ has a qualitative similarity with the \EAI and
the SK model. We computed averages and quantiles for \emph{PK}$0$ and
\emph{PK}$8$ from $10000$ different disorder samples.

\begin{figure}[t]
\begin{center}
\includegraphics[width=0.85\columnwidth]{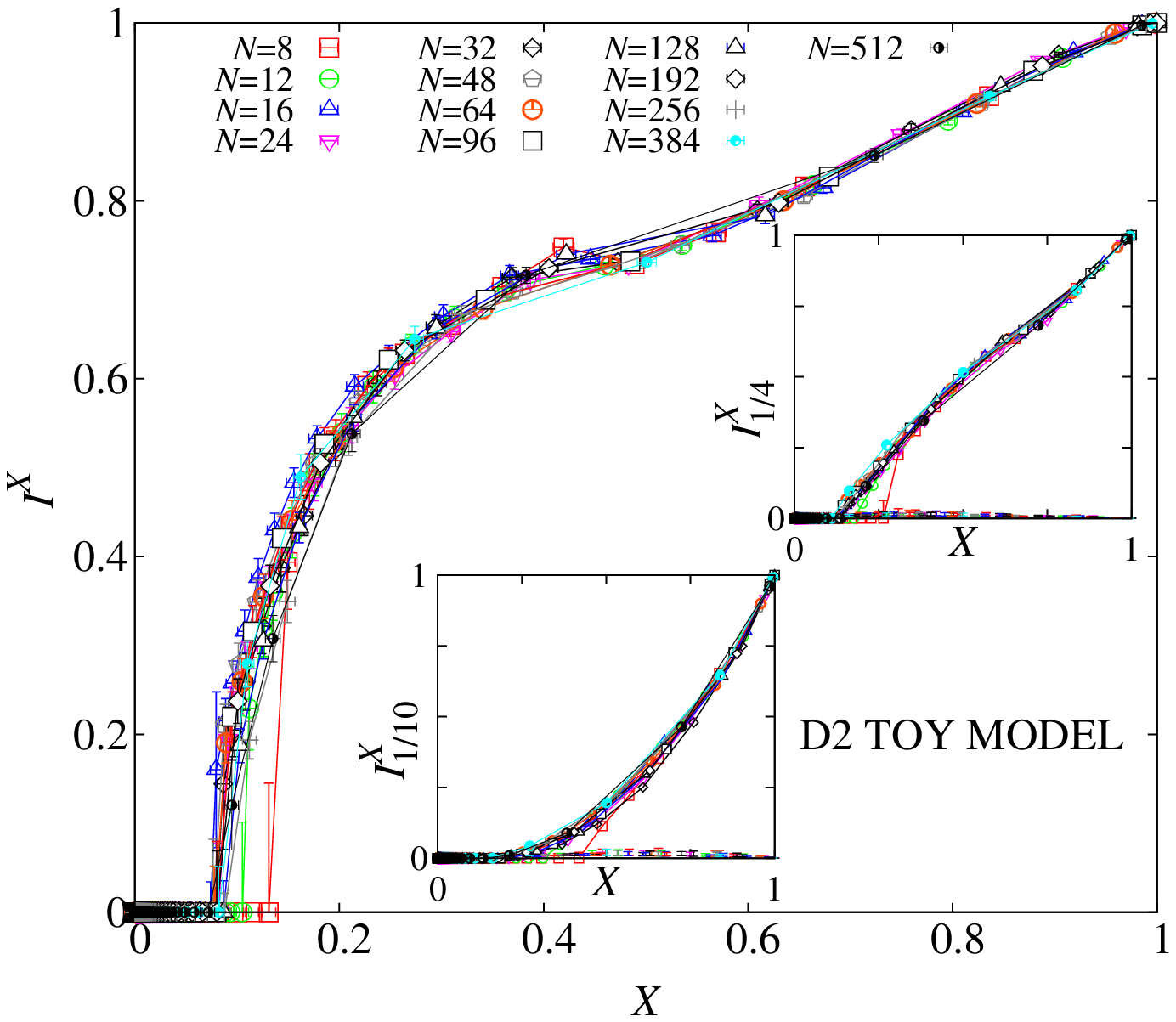}\\
\includegraphics[width=0.85\columnwidth]{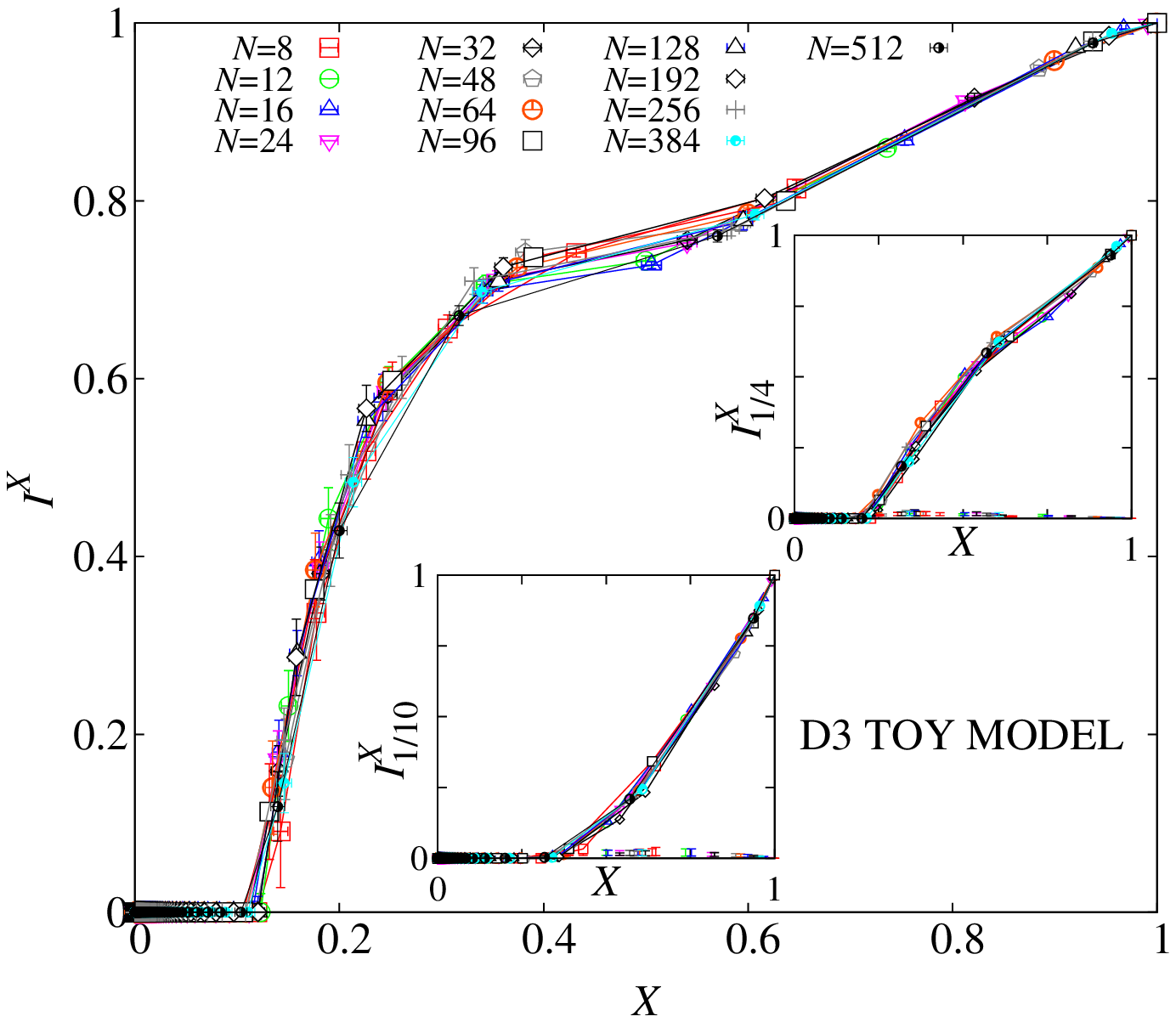}
\caption{(Color online) Droplet-like toy models: $f_p$ as a
  function of $X$, $p=1/2$ (main plots), $p={1/4}$ and $p={1/10}$
  (insets). Top: \emph{D}$2$ toy model. Bottom: \emph{D}$3$.}
\label{fig:droplet}
\end{center}
\end{figure}

We find a completely different behavior in the \emph{D}$2$ and
\emph{D}$3$ models (see Fig.~\ref{fig:droplet}). In this case we
computed overlaps by randomly flipping clusters: to minimize the
effects of a limited number of measurements, we preferred to simulate
a reasonable but not huge number of samples ($1000$) and to collect a
fairly large number of measurements per sample for the largest sizes
($10^{10}$ for $L\geq64$ in \emph{D}$2$ and $10^8$ for $L\geq16$ in
\emph{D}$3$). Although this model has been used
in Ref.~\onlinecite{middleton:13} to provide an example of finite-size effects
persisting up to very large system sizes, the almost perfect collapse
of data for all simulated sizes, when plotted as a function of $X$,
is striking.  The quantities $f_p$ are rapidly decreasing with
$X$, and are almost zero in a wide interval down to $X=0$. The
possibility of finite-statistics effects driving the sudden drop in
$f$ cannot be completely ruled out.
\begin{figure}[t]
\begin{center}
\includegraphics[width=0.85\columnwidth]{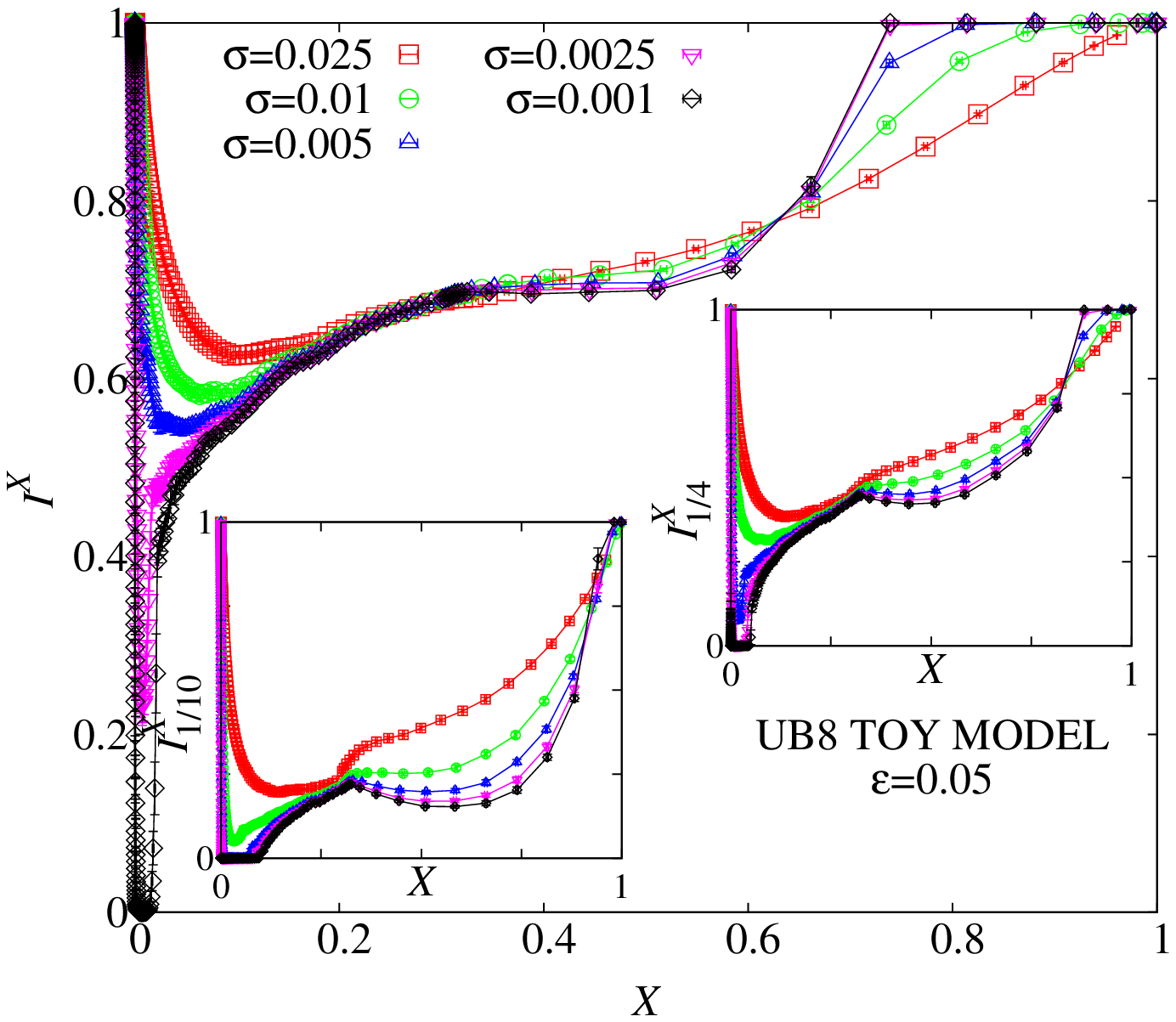}\\
\includegraphics[width=0.85\columnwidth]{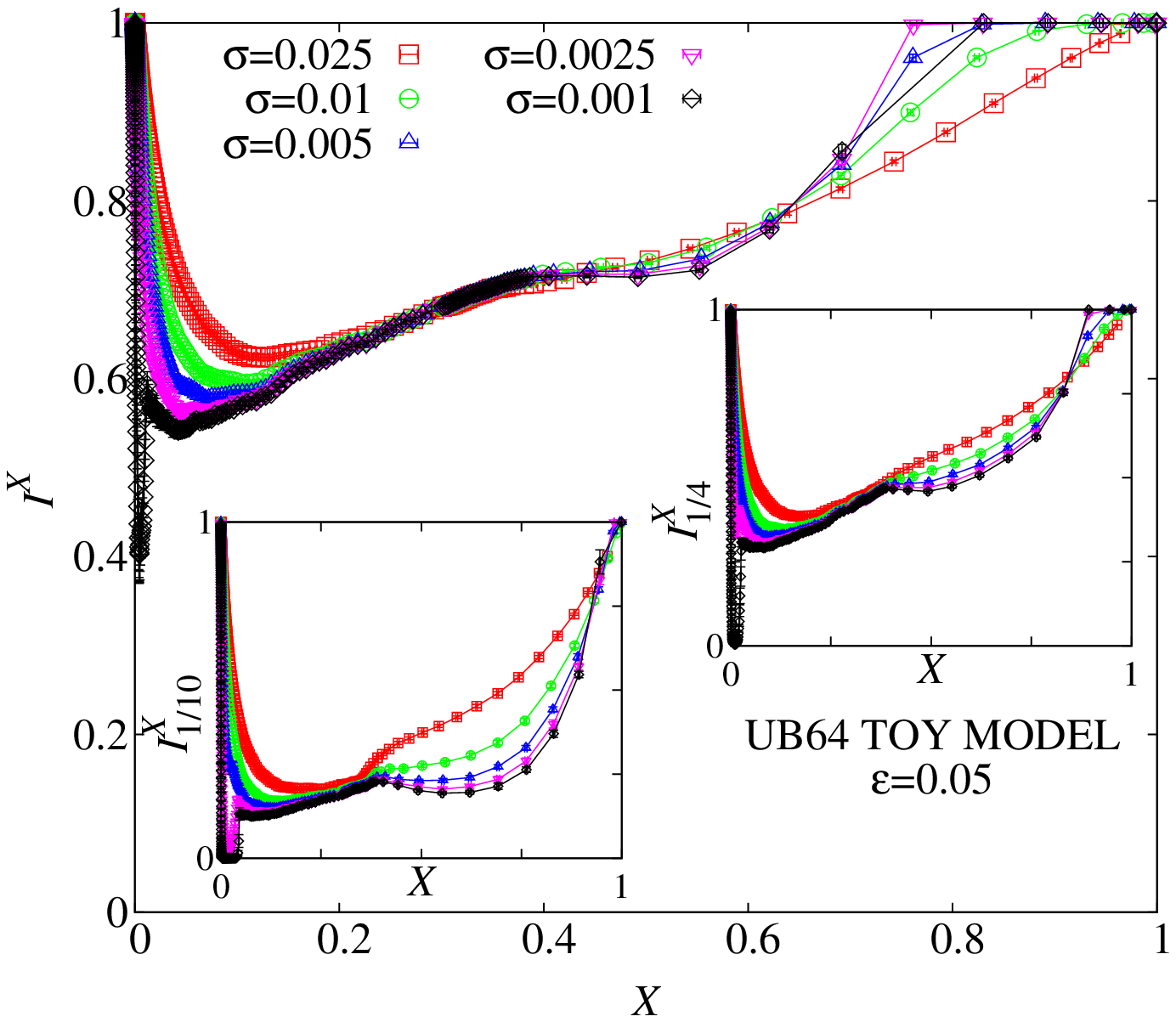}
\caption{(Color on-line) Random bifurcation toy model: $f_p$ as a
  function of X, $p=1/2$ (main plots), $p={1/4}$ and $p={1/10}$
  (insets). Top: \emph{UB}$8$; bottom: \emph{UB}$64$.}
\label{fig:ub}
\end{center}
\end{figure}

Finally, in Fig.~\ref{fig:ub}, we show $f_p$ for $p=1/2$, $1/4$,
$1/10$, for the \emph{UB}$\lambda$ toy model. We have averaged over
$10000$ instances of the quenched noise.  The data for high branching
probability closely resemble those of the SK model, also for the
dependence on the system size (we mimic finite-size effects by tuning
$\sigma$). Since a finite fraction of samples have no weight at low
$X$, a narrow dip is present near $X=0$ for small $\sigma$ value. As
the forking probability grows, so does the fraction of samples with peaks
in the $P_J(q)$ at small $q$ values, and the dip shrinks
away. The curves for larger values of $\sigma$ show the steep rise
towards the $f(X=0)=1$ singularity, as in the case of our data for the
spin-glass models. At small $\sigma$ and large $\lambda$ the curves
have the expected limit $f_p(X{\rightarrow 0})=p$.  Unfortunately, the
spin-glass data do not allow a fair extrapolation of a possible
size-dependent $f_p(0)$ limit to compare with.

The conclusion of this exercise based on toy models is that in order
to produce a behavior of $f_p$ similar to the one observed for the SK
and 3D \EAI models, one needs many states. In particular, the
droplet toy model completely fails to reproduce the observed
qualitative behavior.

\section{CONCLUSIONS}
\label{sec:CONC}

The question of the large-volume extrapolation of numerical data for
the overlap distribution of spin-glass models has been a subject of
controversy over the years. Recently, Ref.~\onlinecite{middleton:13}
has proposed the use of $I(q)$, the median over disorder samples of the
overlap cumulative distribution, showing that for some droplet-like
models it converges rapidly to zero in a whole interval of overlap
values close to the origin. This is very different, and more
clarifying, than the slow convergence of the mean $X(q)$. We use
$I(q)$ (and its generalization to different quantiles) to study the SK
and the $3D$ \EAI models.  The results of the two models are very
similar and unmistakably different from the one that is obtained for
the droplet-like models of Ref.~\onlinecite{middleton:13}, making the
case for a RSB-like behavior of the $3D$ \EAI model in the spin-glass
phase.

We have studied $I_p(q)$, the quantiles over disorder samples of the
overlap cumulative distribution, comparing our numerical estimates of
$f_p(q)=I_p(q)^{X(q)}$ with the predictions of the RSB theory for the
SK model.  The numerical results for the SK model converge (although
non-uniformly) as $N$ grows towards the RSB predictions for the low
$X(q)$ behavior of $f_p(q)$. The results for the \EAI model are again
qualitatively very similar to the one obtained in the mean field
theory, even if the infinite-volume limit seems clearly more difficult
to reach in the finite dimensional theory.

We have also studied several toy models, which show that the observed
behavior of $f_p$ is connected to the existence of many thermodynamic
states.

\begin{acknowledgments}

We thank the Janus Collaboration for allowing us to use their \EAI
data.  The research leading to these results has received funding from
the European Research Council under the European Union's Seventh
Framework Programme (FP7/2007-2013), ERC grant agreement 247328.
V. M.-M. and D. Y. acknowledge support from MINECO (Spain), contract
no. FIS2012-35719-C02.

\end{acknowledgments}

\end{document}